\documentclass[%
 reprint,
 amsmath,amssymb,
 aps,
prb,
]{revtex4-2}
\usepackage{braket}
\usepackage{comment}

\usepackage{graphicx}
\usepackage{tipa}
\usepackage{amsfonts}
\usepackage{bbding}
\usepackage{bbold}
\usepackage{bbm}
\usepackage[T1]{fontenc}
\usepackage{amsmath}
\usepackage{color}
\usepackage{soul}
\usepackage{ulem}
\usepackage{bbm} 
\usepackage{tensor}
\usepackage{dsfont}
\usepackage{slashed}
\usepackage{hyperref}
\usepackage[all]{hypcap}
\hypersetup{colorlinks=true,
   linkcolor=blue,
    citecolor=blue,
    filecolor=magenta,
    urlcolor=cyan}

\begin{document}

\title{ When Adiabaticity Is Not Enough to Study Topological Phases in Solid-State Physics: Comparing the Berry and Aharonov-Anandan Phases in 2D Materials}

\author{Abdiel de Jes\'us Espinosa-Champo$^{a,c}$, Alejandro Kunold $^{b}$ and Gerardo G. Naumis$^{d}$ }
\affiliation{${}^{a}$Posgrado de Ciencias F\'isicas, Universidad Nacional Aut\'onoma de M\'exico, Apartado Postal 20-364 01000, Ciudad de M\'exico, M\'exico.}
\affiliation{${}^{b}$ \'Area de F\'isica Te\'orica y Materia Condensada, Universidad Aut\'onoma Metropolitana Azcapotzalco, Av. San Pablo 180, Col. Reynosa-Tamaulipas, 02200 Cuidad de M\'exico, M\'exico}
\affiliation{${}^{c}$Departamento de F\'isica, Facultad de Ciencias, Universidad Nacional Aut\'onoma de M\'exico, Apdo. Postal 70-542, 04510, CDMX, M\'exico.}
\affiliation{${}^{d}$Depto. de Sistemas Complejos, Instituto de F\'isica, Universidad Nacional Aut\'onoma de M\'exico (UNAM). Apdo. Postal 20-364, 01000, CDMX, M\'exico.}
\email[e-mail: ]{naumis@fisica.unam.mx }
\begin{abstract}

{\color{black}Topological phases emerge as the parameters of a quantum system vary with time. Under the adiabatic approximation, the time dependence can be eliminated, allowing the Berry topological phase to be obtained from a closed trajectory in parameter space. In solid-state physics, this approach is commonly applied by taking a reciprocal space wavevector as the parameter, which is assumed to be varied by electromagnetic fields. The Berry curvature is then obtained by computing the derivatives of Bloch wavefunctions in reciprocal space. However, in many systems—especially gapless ones—the adiabatic approximation is never satisfied. This is particularly true in Dirac and Weyl materials, where the Berry curvature is often calculated without considering the breakdown of the adiabatic condition. In this work, we demonstrate how other time-dependent topological quantities, specifically the Aharonov-Anandan phase, can be used to extract information not only about topology but also about band transitions in 2D materials. In particular, a relationship between the current and the Aharonov-Anandan phase is proved,  showing that photon-induced transitions produce current vortices. To illustrate this, we analyze graphene under electromagnetic radiation from a time-driven perspective, showing how the Aharonov-Anandan and Berry phases provide complementary insights into topology, interband transitions, and currents.  This is achieved by using the Dirac-Bloch formalism and by solving the time-dependent equations within Floquet theory. }

\end{abstract}

\maketitle

\section{Introduction} \label{sec: Introduction}

Beyond electronic properties, the study of quantum geometry and topology in solid-state physics—especially in two-dimensional (2D) materials—has emerged as a pivotal area of research \cite{Bernevig+2013}. Quantum geometry, characterized by parameters such as the Berry curvature and quantum metric, plays a crucial role in understanding various physical phenomena, including the anomalous Hall effect and topological insulators \cite{Wu2016}. Topology, on the other hand, concerns the global properties of the material's electronic wavefunctions, leading to robust edge states and exotic phases of matter \cite{McIver2020}.

In solid-state physics, topology is usually studied  considering the reciprocal space properties of the Hamiltonian and its corresponding wave functions \cite{phillips2012advanced}. However, it is often forgotten that this procedure is valid whenever the parameters of the Hamiltonian are varied adiabatically in time in such a way that there are no transitions to other states. {\color{black} This requires driving the system with a frequency smaller than the gap size measured in frequency units}. It turns out that in many cases the adiabatic condition does not hold as we will discuss here\cite{Wang:16}. In fact, many 2D materials fall into this category. Among these, graphene and borophene stand out. Graphene has been widely celebrated for its exceptional conductivity and mechanical strength \cite{Oka2009}, and borophene for its remarkable flexibility and electronic characteristics \cite{Basov2017}. 

Although numerous studies explore the topological properties of these 2D materials, it is often overlooked that many lack a band gap, and thus, the adiabaticity condition is never truly met. Some researchers gap the system's spectrum by breaking a symmetry, yet this does not answer what the required size of the gap should be or whether this process changes the nature of the topology. 

{ \color{black}Since the adiabaticity condition does not hold, the alternative is to focus on the time-dependent origin of topological phases, as done in Thouless's pioneering work on topology \cite{PhysRevLett.49.405}.}
Fortunately, in other areas of physics, different dynamic approaches have been developed to identify topological properties \cite{cohen2019geometric,Zhu_2022}. {\color{black} The aim of the present work is to compare the information provided by the Aharonov-Anandan phase with the Berry phase in the context of solid-state band topology. The advantage is that the Aharonov-Anandan phase does not require adiabaticity \cite{Zhu_2022}.  As we shall see, in many instances, the Aharonov-Anandan and Berry phases are radically different and encode distinct physical aspects of the electronic properties of 2D materials. 
Until recently, the consensus was that the Berry and Aharonov-Anandan phases coincide in the adiabatic limit \cite{PhysRevLett.58.1593,ZHAOYANWU1996201}.
However, it was recently discovered that degeneracy can cause them to differ \cite{Zhu_2022}. Our work aligns with this perspective.
In solid-state physics, degeneracy and gapless systems are common, necessitating a careful examination of the equivalence between topological phases.} 

The understanding of such time-dependent aspects opens new avenues in material science and condensed matter, as it allows us to manipulate the properties of these materials through external stimuli 
\cite{bukov2015universal,Chan2017Photocurrents, Champo2021}. One such method is time-driving, which consists in the temporal modulation of a material's properties using external fields. This technique not only facilitates the exploration of new phases of matter but also offers a platform for the dynamic control of electronic and topological features \cite{Calvo2013,Rodriguez2021}.

The application of electromagnetic fields to 2D materials offers a powerful means to alter their quantum geometry and topology. By tuning the frequency, amplitude, and polarization of the applied fields, researchers can induce significant modifications in the band structure and topological properties of the material \cite{Basov2017}. For instance, it is possible to induce topological phase transitions, where the material's topological invariants change, leading to new states of matter with distinct edge modes and transport properties \cite{Wang2013}. These field-induced modifications not only enhance the fundamental understanding of 2D materials but also pave the way for novel applications in electronic and optoelectronic devices \cite{McIver2020}.

The ability to dynamically control the properties of 2D materials through time-driving and electromagnetic fields has far-reaching implications for technology \cite{CARVALHO201724}. Potential applications include the development of high-speed, low-power electronic devices and advanced sensors \cite{CARVALHO201724,Carvalho2018}. The tunability provided by these techniques allows the design of materials with tailored properties, optimized for specific applications \cite{Wu2016}. For example, the induction of topological states can lead to devices with robust edge conduction, immune to scattering and defects, which is highly desirable for reliable electronic and spintronics applications \cite{Rudner2013}.

Time-driving in 2D materials typically involves the application of periodic external fields, such as electromagnetic or acoustical waves, which induce a time-dependent perturbation in the system \cite{Babek_2014,Babek_2016,Abdiel2019,Ibarra-Sierra_2022}. The response of these materials to this driving can be effectively studied using the Floquet theory \cite{Wang2013,Babek_2014,Babek_2019}. Floquet theory extends the concept of Bloch's theorem to time-periodic systems, allowing the analysis of the system's behavior in terms of quasi-energy states known as Floquet states. These states provide a comprehensive understanding of the material's dynamic response, enabling the identification of novel phenomena such as Floquet-Bloch bands and time-crystalline states \cite{Rudner2013,Abanin_2023}. 

{\color{black} In summary, in this article we provide an alternative path, based on Floquet theory, to obtain topological phases for solid-state systems that are gapless. As a result, we relate  non-adiabatic topological phases with electronic transitions and currents induced by the applied electromagnetic field.} To do so, we will compare and explore the Berry and Aharonov-Anandan phases by the implementation of time-driving in materials such as graphene or borophene, delving into how Floquet theory and Dirac-Bloch equations serve as essential tools for understanding and predicting their properties. The layout of this work is as follows. In Sec. \ref{sec:topological}, we discuss the Berry and Aharonov-Anandan phases from a time-dependent driving perspective.
Section \ref{sec:Numerical Methods} presents the results for graphene, followed by remarks on the structure of the phases in Section \ref{sec:Discussion and conclusions} and induced currents in  Section \ref{Sec:Currents}.  Finally, Section \ref{sec:Conclusions} provides a summary of our findings.

\section{Berry and Aharonov-Anandan topological phases from a time-dependent driving perspective}
\label{sec:topological}

In this section, we review alternative approaches for deriving adiabatic (Berry) and non-adiabatic (Aharonov-Anandan) topological phases from a time-dependent perspective.
We consider the case of a typical 2D Dirac material, graphene. As is well known, the electronic spectrum of such a platform lacks a band gap and instead forms Dirac cones, with each valley $K_{\pm}$ hosting an effective massless electron.
For such a gapless spectrum, the adiabatic condition is never truly satisfied, yet a topological charge is usually assigned to each Dirac point \cite{Carvalho2018,CARVALHO201724}.
Let us discuss this situation from a time-dependent perspective. Consider the Schr\"odinger equation
\begin{equation}
i\hbar \partial_t \Psi(\boldsymbol{r},t)={H}_{\xi}\Psi(\boldsymbol{r},t),
\label{eq:scrhodinger1}
\end{equation}
where the graphene's Hamiltonian is given by
\begin{equation}
{H}_{\xi}=v_F\boldsymbol{\sigma}_{\xi}\cdot\left[{\boldsymbol{p}}
+e\boldsymbol{A}(\boldsymbol{r},t)\right],
\end{equation}
$v_F\approx 10^6$\,m/s , the subscript $\xi=\pm 1$ tags the Dirac valley and $\boldsymbol{\sigma}_{\xi}=(\xi \sigma_x,\sigma_y)$. {\color{black} $\sigma_x$ and $\sigma_y$ are the usual $2\times 2$ Pauli matrices.}
In the special case of normal incidence, the vector potential is independent of position. Eq. \eqref{eq:scrhodinger1} simplifies significantly when we assume the ansatz
\begin{equation}
\Psi(\boldsymbol{r},t)=\exp\left(\boldsymbol{k}\cdot\boldsymbol{r}\right)
\Psi_{\boldsymbol{k}}(t),
\end{equation}
leading to the $\boldsymbol{k}$-space Schrödinger equation:
\begin{equation} 
	i \hbar\frac{\partial}{\partial t}\boldsymbol{\Psi}_{\boldsymbol{k}}(t) 
    = {H}_{\boldsymbol{k},\xi}(t)\boldsymbol{\Psi}_{\boldsymbol{k}}(t),
    \label{eq:mainH}
\end{equation}
where the $\boldsymbol{k}$-space Hamiltonian is
\begin{equation}
{H}_{\boldsymbol{k},\xi}=\hbar v_F\boldsymbol{\sigma}_{\xi}
\cdot \boldsymbol{K}(t).
\label{eq:khamiltonian}
\end{equation}
and
$\boldsymbol{K}(t)=\boldsymbol{k}+e\boldsymbol{A}(t)/\hbar$.
This Hamiltonian accounts for the effects of a homogeneous,
periodic external vector potential $\boldsymbol{A}(t)$
with period $T$, which induces changes in the Hamiltonian parameters.
The components of 
$\mathbf{\Psi}_{\boldsymbol{k}}(t)
=(\Psi_{\boldsymbol{k},A}(t),
\Psi_{\boldsymbol{k},B}(t))$
are interpreted as the amplitudes of the electron wave function
on each of the graphene's bipartite lattices,
denoted here by $A$ and $B$. 
After these definitions are introduced,
the Berry and Aharonov-Anandan approaches are discussed
in the following subsections.

\subsection{Berry phases: the Dirac-Bloch formalism approach}
\label{sec:Dirac-Bloch formalism}
The Berry phase is typically discussed in the context of the
adiabatic approximation 
\cite{Berry1984,Barry1984,cohen2019geometric}.
In this framework,
the system's dynamics are primarily encoded in the Berry and 
dynamic phases, with the system predominantly remaining in a 
single quantum state  without undergoing transitions between 
levels.

Another way to frame this problem, however, is through
the Dirac-Bloch  equations (DBEs). In this approach,
the Berry phase is defined in the same  way as in the
adiabatic approximation, but the adiabatic conditions are not 
required to be satisfied 
\eqref{eq:ansatz_time_dependent_solution}. Instead, 
while the intra-level dynamics are still governed by
the Berry phase, transitions are accounted for through
the so-called Rabi phase.

The Dirac-Bloch equations (DBEs) are pivotal in the study of the 
nonlinear  optical responses in advanced materials 
\cite{Ishikawa2010,DiMauro2022}, 
especially those with complex electronic band structures like 
twisted  bilayer graphene (TBG) \cite{DiMauro2022,Ornigotti2023}.
The DBEs provide a detailed description of the 
interaction between electrons and electromagnetic fields,
incorporating both intraband and interband transitions.
In this section, we focus on this 
powerful method as it allows to analyze the behavior
of the Berry phase.

The solution to the quantum evolution Eq. \eqref{eq:mainH}
can be written as \cite{DiMauro2022},
{\color{black}
\begin{equation} \label{eq:ansatz_time_dependent_solution}
\Psi_{\boldsymbol{k},\xi}(t) = \sum_{\lambda} c^\lambda_{\boldsymbol{k},\xi}(t)
\varphi^\lambda_{\boldsymbol{k},\xi}(t)
\mathrm{e}^{-i \gamma^\lambda_{D,\boldsymbol{k},\xi}(t)},
\end{equation}}
where {\color{black}$ c^\lambda_{\boldsymbol{k},\xi}(t)$} are time-dependent coefficients, $ 
\lambda =\pm 1$ denotes the band index, and 
{\color{black}$\gamma^\lambda_{D,\boldsymbol{k},\xi}(t)$} is the dynamical phase factor. The instantaneous eigenstates
\begin{equation}
\varphi_{\boldsymbol{k},\xi}^{\lambda}
=\frac{1}{\sqrt{2}}\left(1,
\lambda\exp(i \xi\theta_{\boldsymbol{k}}(t)\right)^\intercal
\,\,\,\,, \lambda=\pm 1,
\label{eq:insteigstates}
\end{equation}
with the phase 
\begin{equation}
    \theta_{\boldsymbol{k}}(t)= \tan^{-1}(K_y(t)/K_x(t)),
    \label{eq:berryangle}
\end{equation}
where $K_x(t)$ and $K_y(t)$ are the components
of $\boldsymbol{K}(t)$, satisfy the instantaneous Schr\"odinger equation with instantaneous eigenvalues $\epsilon^\lambda_{\boldsymbol{k}}(t)$,
\begin{equation} \label{eq:instantaneous_time_independent_schrödinger_equation}
{H}_{\boldsymbol{k},\xi}(t)\varphi^\lambda_{\boldsymbol{k},\xi}(t) = \epsilon^\lambda_{\boldsymbol{k}}(t)\varphi^\lambda_{\boldsymbol{k},\xi}(t).
\end{equation}
where
\begin{equation}
\epsilon^\lambda_{\boldsymbol{k}} (t)=\lambda \hbar v_F
\left\vert\boldsymbol{K}(t)\right\vert
\label{eq:instantstate}
\end{equation} 
\textcolor{black}{is independent on the valley index}. The dynamical phase is obtained  from the instantaneous eigenvalues  and is defined as \cite{Villari2018}:
\begin{equation} \label{eq:dynamical_phase_factor_bands}
\gamma^{\lambda}_{D,\boldsymbol{k}}(t) = \frac{1}{\hbar}\int_{0}^t \epsilon^\lambda_{\boldsymbol{k}}(t')\, dt'.
\end{equation}

{\color{black}    
The Berry phase $\gamma^\lambda_{B,\boldsymbol{k},\xi}(t)$ 
can now be extracted from \cite{Villari2018},
\begin{multline} 
\gamma^\lambda_{B,\boldsymbol{k},\xi}(t) 
= i \int_{0}^t  \varphi^\lambda_{\boldsymbol{k},\xi}(t')
\frac{d \varphi^\lambda_{\boldsymbol{k},\xi}(t')}{dt'}    dt',\\
=  i \xi \int_{0}^t  \varphi^\lambda_{\boldsymbol{k},|\xi|}(t')
\frac{d \varphi^\lambda_{\boldsymbol{k},|\xi|}(t')}{dt'}
dt',\\
\equiv \xi \gamma^\lambda_{B,\boldsymbol{k}}(t).
\label{eq:Berry_phase_definition}
\end{multline}
}
\textcolor{black}{Therefore, the Berry phases in the two valleys
have opposite signs.}
When the time in integral \eqref{eq:Berry_phase_definition} is 
evaluated from $t=0$ to $t=T$, the trajectory followed by $
\boldsymbol{K}(t) $  
in the $\boldsymbol{k}$-parameter space defines a contour $
\mathcal{C}_B$, which may or may not encircle the origin $n \in 
\mathbb{Q}$ times per period. Thus, the geometrical Berry phase 
becomes
{\color{black}
\begin{multline} \label{eq:Geometrical-Berry-phase}
    \gamma_{B, \boldsymbol{k},\xi}^{\lambda} \equiv \gamma_{B, \boldsymbol{k},\xi}^{\lambda} (T)\\
    =i \oint_{\mathcal{C}_{B}}  \varphi^{-\lambda}_{\boldsymbol{k},\xi}(t') \nabla_{\boldsymbol{K}(t')} \varphi^\lambda_{\boldsymbol{k},\xi}(t') d \boldsymbol{K}(t')\\
     =\xi i \oint_{\mathcal{C}_{B}}  \varphi^{-\lambda}_{\boldsymbol{k},|\xi|}(t') \nabla_{\boldsymbol{K}(t')} \varphi^\lambda_{\boldsymbol{k},|\xi|}(t') d \boldsymbol{K}(t')=  \xi n \pi.
\end{multline}}
Although the Berry phase is only well-defined in the adiabatic 
case, as discussed in Appendix \ref{Appendix: Approximation-Time-Evolution}, it contributes to the total geometric phase in the 
non-adiabatic case.

Substituting this proposed solution into the time-dependent Schr\"odinger equation and simplifying, we obtain the time evolution of the coefficients:
{\color{black}
\begin{multline} 
\frac{dc^\lambda_{\boldsymbol{k},\xi}(t)}{dt} 
= i \Omega_{\boldsymbol{k},\xi}(t) c^{-\lambda}_{\boldsymbol{k},\xi}(t)
\exp\left[
  i\gamma_{D,\boldsymbol{k}}^{\lambda}(t)
- i\gamma_{D,\boldsymbol{k}}^{-\lambda}(t) \right]\\
 \times\exp\left[
  i\gamma^\lambda_{B,\boldsymbol{k},\xi}(t)
- i\gamma^{-\lambda}_{B,\boldsymbol{k},\xi}(t))\right],
\label{eq:differential_equations_for_coefficients}
\end{multline}
}
where we have introduced the Rabi phase
{\color{black}
\begin{equation} \label{eq:Rabi_dynamical_phase}
\gamma_{R,\boldsymbol{k},\xi}(t)= \int_{0}^{t} dt' \nu_{\boldsymbol{k},\xi}(t').
\end{equation}}
and the Rabi frequency {\color{black}$\nu_{\boldsymbol{k},\xi}(t)$} \cite{DiMauro2022},
{\color{black}
\begin{multline}
\nu_{\boldsymbol{k},\xi}(t)
=i \varphi^{-\lambda}_{\boldsymbol{k},\xi}(t) 
\frac{d\varphi^\lambda_{\boldsymbol{k},\xi}(t)}{dt}\\
={\xi} i  \varphi^{-\lambda}_{\boldsymbol{k},|\xi|}(t) 
\frac{d\varphi^\lambda_{\boldsymbol{k},|\xi|}(t)}{dt}
\equiv {\xi} \nu_{\boldsymbol{k}}(t).
\label{eq:Rabi-frequency-definition}
\end{multline}
}

\subsection{Aharonov-Anandan phases: the Floquet formalism approach}
\label{sec:Aharanovformalism}

When the Hamiltonian is periodic with period $T=2\pi/\Omega$,
due to the external 
driving, the solution to Eq. (\ref{eq:mainH}) can be
obtained using the Floquet theorem.
In this case, it is convenient to rewrite
Schr\"odinger's equation
\eqref{eq:mainH}
for the time evolution operator as
\begin{equation}
i\hbar \frac{d}{dt} U_{\boldsymbol{k},\xi}(t)
=H_{\boldsymbol{k},\xi}(t)U_{\boldsymbol{k},\xi}(t).
\label{eq:schroU}
\end{equation}
The Floquet theorem states that the time evolution operator
must take the form \cite{Dai2016,Ibarra-Sierra_2022}
\begin{equation}
U_{\boldsymbol{k},\xi}(t) = \exp\left(-\frac{i}
{\hbar}{H}_{e,\boldsymbol{k},\xi} \,\, t\right)
{W}_{\boldsymbol{k},\xi}(t), \label{eq:UFloquet}
\end{equation}
where ${H}_{e,\boldsymbol{k},\xi}$ is a time-independent matrix known as the 
effective Hamiltonian. Its eigenvalues are the quasienergies
given by
\begin{equation}
\mathcal{E}_{\boldsymbol{k},\xi}^j = -\frac{\hbar \Omega}{2\pi} 
\arg(u_{\boldsymbol{k},\xi}^j) + n \hbar \Omega,
\end{equation}
where $u_{\boldsymbol{k},\xi}^j$ is the $j$-th eigenvalue
of $U_{\boldsymbol{k},\xi}(T)
=\exp(-i{H}_{e,\boldsymbol{k},\xi} T/\hbar)$. The corresponding
eigenvectors are $ \boldsymbol{\Psi}_{\boldsymbol{k},\xi}^j(0)$
and the time-dependent Floquet states are expressed as
\begin{multline}
\boldsymbol{\Psi}_{\boldsymbol{k},\xi}^j(t) = \exp\left(-\frac{i}
{\hbar}{H}_{e,\boldsymbol{k},\xi} t\right) {W}_{\boldsymbol{k},\xi}(t) 
\boldsymbol{\Psi}_{\boldsymbol{k},\xi}^j(0)\\
= \exp\left(-\frac{i}{\hbar}\mathcal{E}_{\boldsymbol{k},\xi}^j t\right)
\boldsymbol{P}_{\boldsymbol{k},\xi}(t),
 \label{eq:Floquet}
\end{multline}
where the matrix ${W}_{\boldsymbol{k},\xi}(t)$, known as the monodromy matrix,
is periodic, ${W}_{\boldsymbol{k},\xi}(t)={W}_{\boldsymbol{k},\xi}(t+T)$,
and in particular ${W}_{\boldsymbol{k},\xi}(T)=1$. The spinor
$\boldsymbol{P}_{\boldsymbol{k},\xi}(t+T)=\boldsymbol{P}_{\boldsymbol{k},\xi}(t)={W}_{\boldsymbol{k},\xi}(t)\boldsymbol{\Psi}_{\boldsymbol{k},\xi}^j(0)$ is also periodic. 
Notice here that the set of Floquet eigenvectors ${\Psi}_{\boldsymbol{k},\xi}^j(0)$ 
do not necessarily coincide with the initial wave function ${\Psi}_{\boldsymbol{k},\xi}(0)$ written in the basis of the conduction and valence bands. 

In close analogy to the Berry phase, the Aharonov-Anandan
phase is related to the dynamical phase, which is given by
\begin{equation}
\gamma_{D,\boldsymbol{k},\xi}^j(T)
=\frac{1}{\hbar}\int_0^T
\boldsymbol{\Psi}_{\boldsymbol{k},\xi}^{j\dagger}(t)
{H}_{\boldsymbol{k},\xi}
\boldsymbol{\Psi}_{\boldsymbol{k},\xi}^j(t) \,\, dt.
\end{equation}
Consequently, the non-adiabatic Aharonov-Anandan phase can be 
obtained from the Floquet wave function after one period $T$ by 
subtracting the dynamical phase \cite{Zhu_2022}:
\begin{equation}
\boldsymbol{\Psi}_{\boldsymbol{k},\xi}^j(T)
= \mathrm{e}^{-i\gamma_{D,\boldsymbol{k},\xi}^j(T)
-i\gamma_{A,\boldsymbol{k},\xi}^j(T)}
\boldsymbol{\Psi}_{\boldsymbol{k},\xi}^j(0).
\end{equation}
In turn, from Eq. \eqref{eq:Floquet},
the quasienergies, as well as the dynamical
and Aharonov-Anandan phases, are connected through 
\cite{Page_1987,Moore_1990}
\begin{equation}
\frac{\mathcal{E}_{\boldsymbol{k},\xi}^jT}{\hbar}
=\gamma_{D,\boldsymbol{k},\xi}^j(T)+\gamma_{A,\boldsymbol{k},\xi}^j(T).
\label{eq:aharonovphases}
\end{equation}

In this particular case, the non-adiabatic Aharonov-Anandan phase can be extracted from Eq. (\ref{eq:Floquet}), giving \cite{Page_1987}:
\begin{equation}
\gamma_{A,\boldsymbol{k},\xi}^j(t)=\frac{i}{2}\int_0^t
  \frac{w_{\boldsymbol{k},\xi}^{j*}(t) d w_{\boldsymbol{k},\xi}^j(t)
  -w_{\boldsymbol{k},\xi}^j(t) d w^{j*}_{\boldsymbol{k},\xi}(t)}
  {1+\left\vert w_{\boldsymbol{k},\xi}^j(t)\right\vert^2}
  \label{eq:anandan}
\end{equation}
where
\begin{multline}
w_{\boldsymbol{k},\xi}^j(t)
= \frac{\Psi_{\boldsymbol{k},\xi,2}^j(t)}
{\Psi_{\boldsymbol{k},\xi,1}^j(t)}=\left| \frac{\Psi_{\boldsymbol{k},\xi,2}^j(t)}
{\Psi_{\boldsymbol{k},\xi,1}^j(t)} \right| \frac{e^{i\xi \Theta_{\boldsymbol{k},2}(t)}}{e^{i\xi \Theta_{\boldsymbol{k},1}(t)}}\\
=\left| \frac{\Psi_{\boldsymbol{k},\xi,2}^j(t)}
{\Psi_{\boldsymbol{k},\xi,1}^j(t)} \right|  e^{i\xi \Theta_{\boldsymbol{k}}(t)},
\label{eq:wdef}
\end{multline}
and $\Psi_{\boldsymbol{k},\xi,1}^j(t)$ and $\Psi_{\boldsymbol{k},\xi,2}^j(t)$ are the first and second components of the spinor
$\boldsymbol{\Psi}{\boldsymbol{k},\xi}^j(t)$,
respectively, with corresponding instantaneous phases
$\Theta_{\boldsymbol{k},j}(t)$.

Even though the Dirac-Bloch and Floquet formalisms yield
phases with quite strong similarities, the phases in general are radically different and encode different physical aspects of the system.
For example, while the Dirac-Bloch formalism expresses phases in terms of instantaneous solutions, the Floquet formalism requires the use of Floquet states to ensure that the wave function trajectories form closed loops.
Most importantly, the phases arising from the Dirac-Bloch formalism are entirely determined by the contour of the vector potential.
In contrast, the phases described by the Floquet formalism depend solely on the contour of the wave function in complex space.

When the time integral \eqref{eq:anandan}
is performed from $t=0$ to $t=T$ we can
assume that the integration follows the geometric
contour $C_A$ in the complex plane of the variable 
$w_{\boldsymbol{k},\xi}^j(t)$.
Moreover, due to
the Floquet theorem that also states that
$\boldsymbol{\Psi}_{\boldsymbol{k},\xi}^j(t)
=\exp(-i\mathcal{E}_{\boldsymbol{k},\xi}^jt/\hbar)
U_{\boldsymbol{k},\xi}^j(t)$ where
$U_{\boldsymbol{k},\xi}^j(t+T)=U_{\boldsymbol{k},\xi}^j(t)$,
we can ensure that the contour closes.
Consequently, the Aharonov-Anandan phase becomes
\begin{multline}
\gamma_{A,\boldsymbol{k},\xi}^j(T)=\frac{i}{2}\oint_{C_A}
  \frac{w_{\boldsymbol{k},\xi}^{j*} d w_{\boldsymbol{k},\xi}^j
  -w_{\boldsymbol{k},\xi}^j d w^{j*}_{\boldsymbol{k},\xi}}
  {1+\left\vert w_{\boldsymbol{k},\xi}^j\right\vert^2},\\
  =\xi \frac{i}{2}\oint_{C_A}
  \frac{w_{\boldsymbol{k},|\xi|}^{j*} d w_{\boldsymbol{k},|\xi|}^j
  -w_{\boldsymbol{k},|\xi|}^j d w^{j*}_{\boldsymbol{k},|\xi|}}
  {1+\left\vert w_{\boldsymbol{k},|\xi|}^j\right\vert^2}\\
  \equiv \xi \gamma_{A,\boldsymbol{k}}^j(T),
  \label{eq:anandangeometric}
\end{multline}
acquiring a geometrical character similar to the Berry phase.
However, while the integration contour $C_B$ of the Berry phase
is defined by the trajectory of $\boldsymbol{A}(t)$,
the contour of the Aharonov-Anandan phase is determined by the trajectory of the wave function, which can only be obtained by solving the Schrödinger equation.

In the next section, we determine the phases arising from the
two formalisms and compare them under different electromagnetic
field parameters with ${\color{black}{\xi}=+1}$.

\begin{figure}
\centering
\large{a)}
\includegraphics[width=0.9\linewidth]{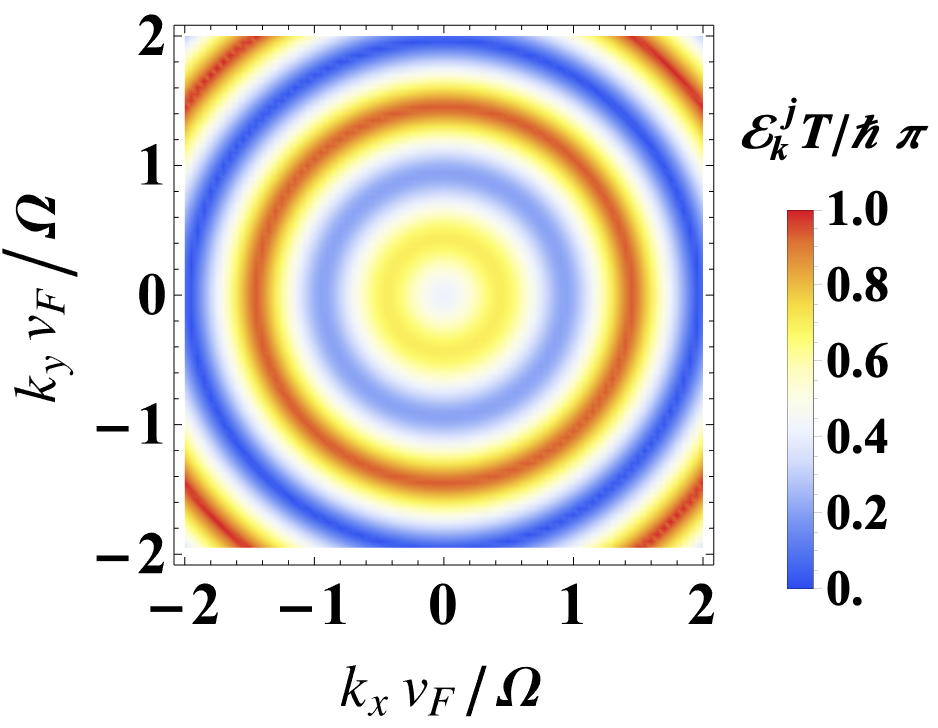}
\large{b)}
\includegraphics[width=0.9\linewidth]{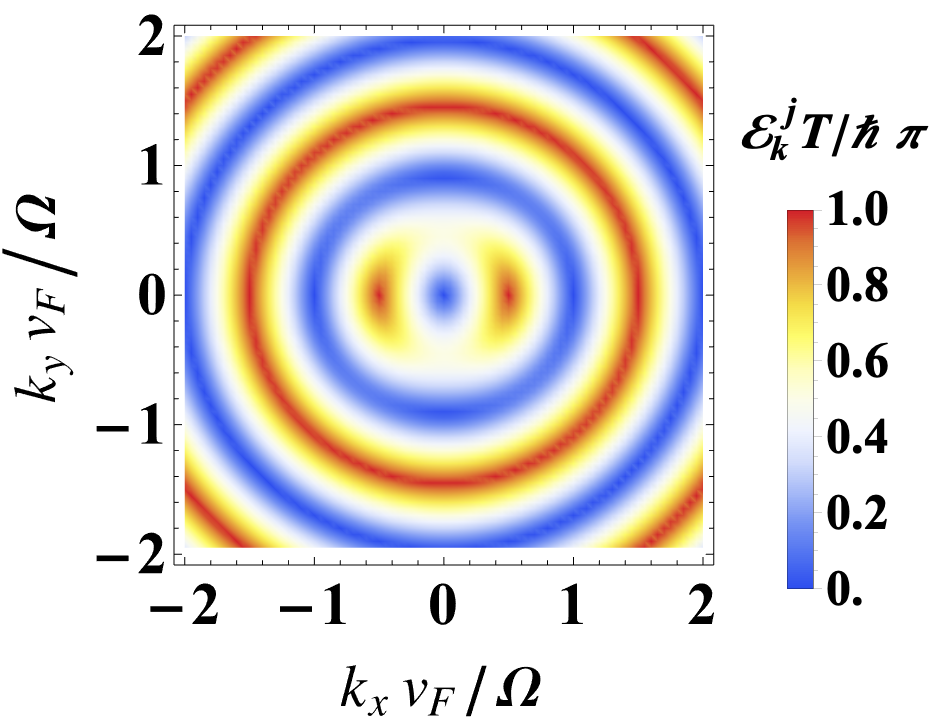}
\caption{Quasi energy spectrum
$\mathcal{E}_{\boldsymbol{k}}^jT/\hbar$
as a function of
$k_x$ and $k_y$ in the weak electric fields regime for (a) circularly polarized light
($ev_FE_x/\hbar \Omega^2=0.5$, $ev_FE_y/\hbar \Omega^2=0.5$)
and (b) linearly polarized light
($ev_FE_x/\hbar \Omega^2=0.5$, $ev_FE_y/\hbar \Omega^2=0.0$). }
\label{fig:spectrumlowfield}
\end{figure}

\begin{figure}
\centering
\large{a)}
\includegraphics[width=0.9\linewidth]{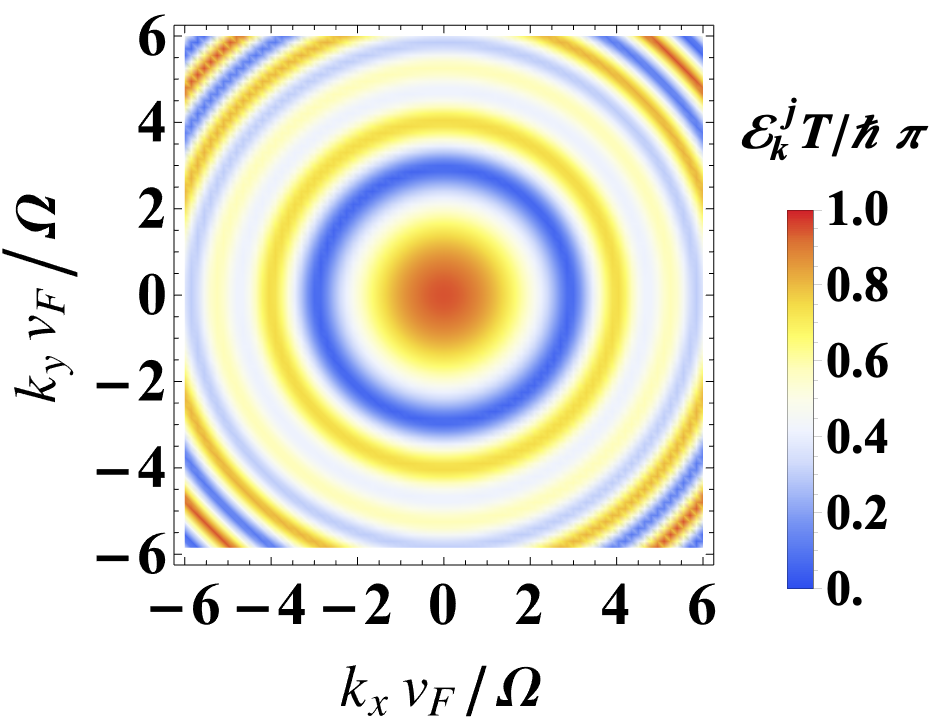}
\large{b)}
\includegraphics[width=0.9\linewidth]{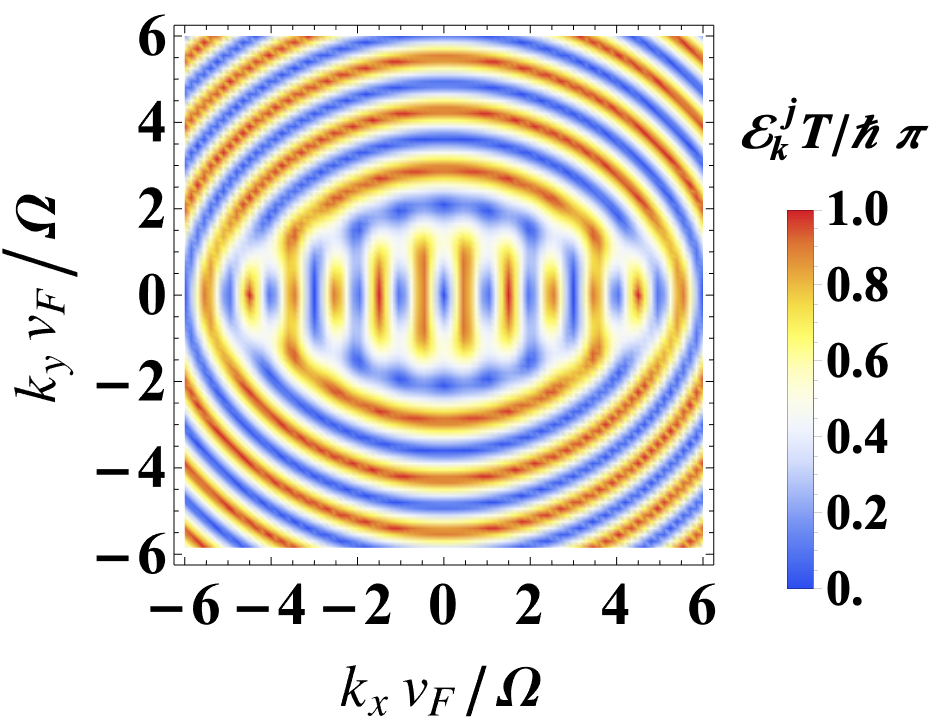}
\caption{Quasi energy spectrum
$\mathcal{E}_{\boldsymbol{k}}^jT/\hbar$
as a function of
$k_x$ and $k_y$ in the strong electric field regime for (a) circularly polarized light
($ev_FE_x/\hbar \Omega^2=ev_FE_y/\hbar \Omega^2=5.0$)
and (b) linearly polarized light
($ev_FE_x/\hbar \Omega^2=5.0$, $ev_FE_y/\hbar \Omega^2=0.0$).
}
\label{fig:spectrumhighfield}
\end{figure}

\begin{figure}
\centering
\large{a)} \includegraphics[width=0.9\linewidth]{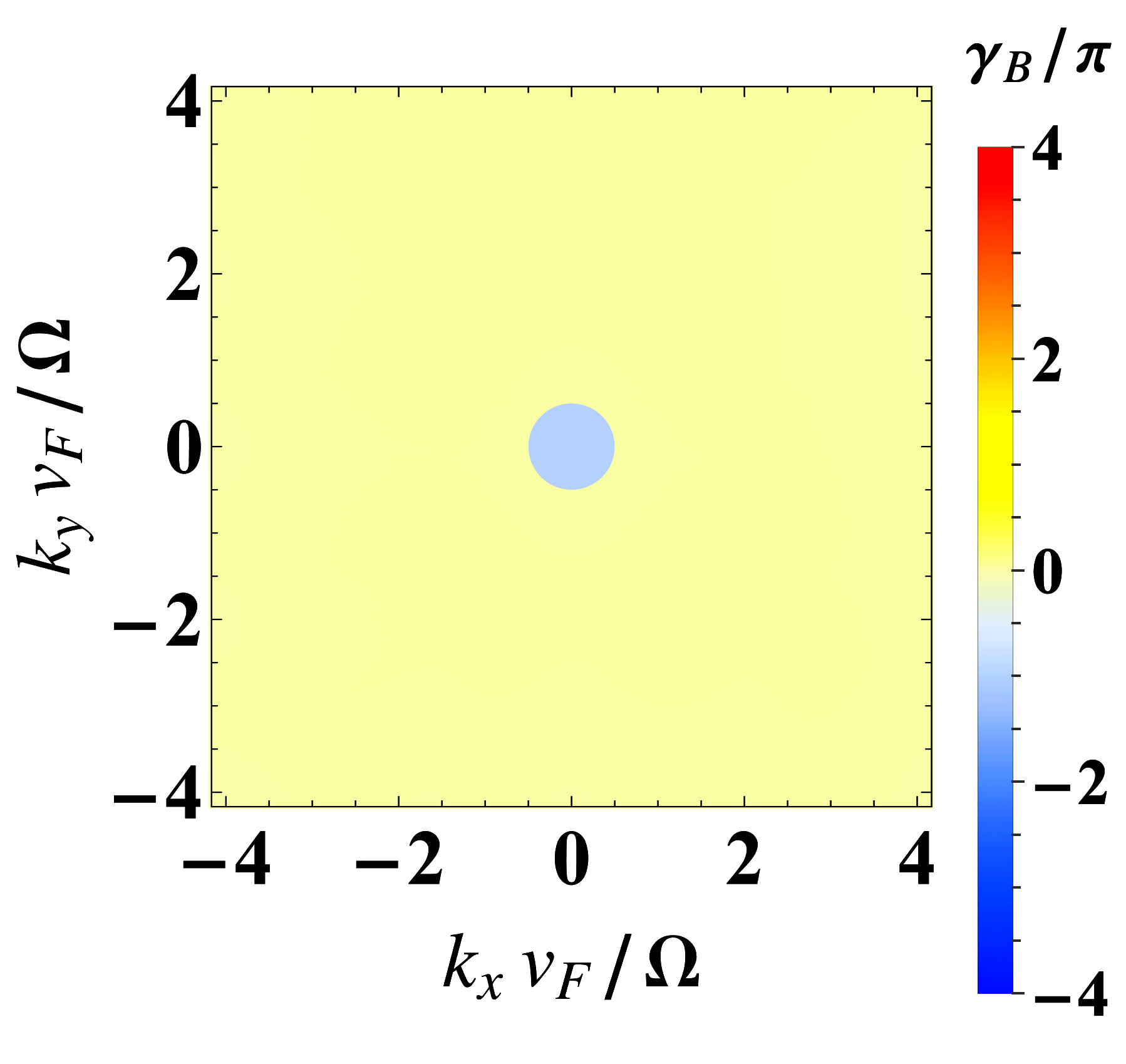}
\large{b)}\includegraphics[width=0.9\linewidth]{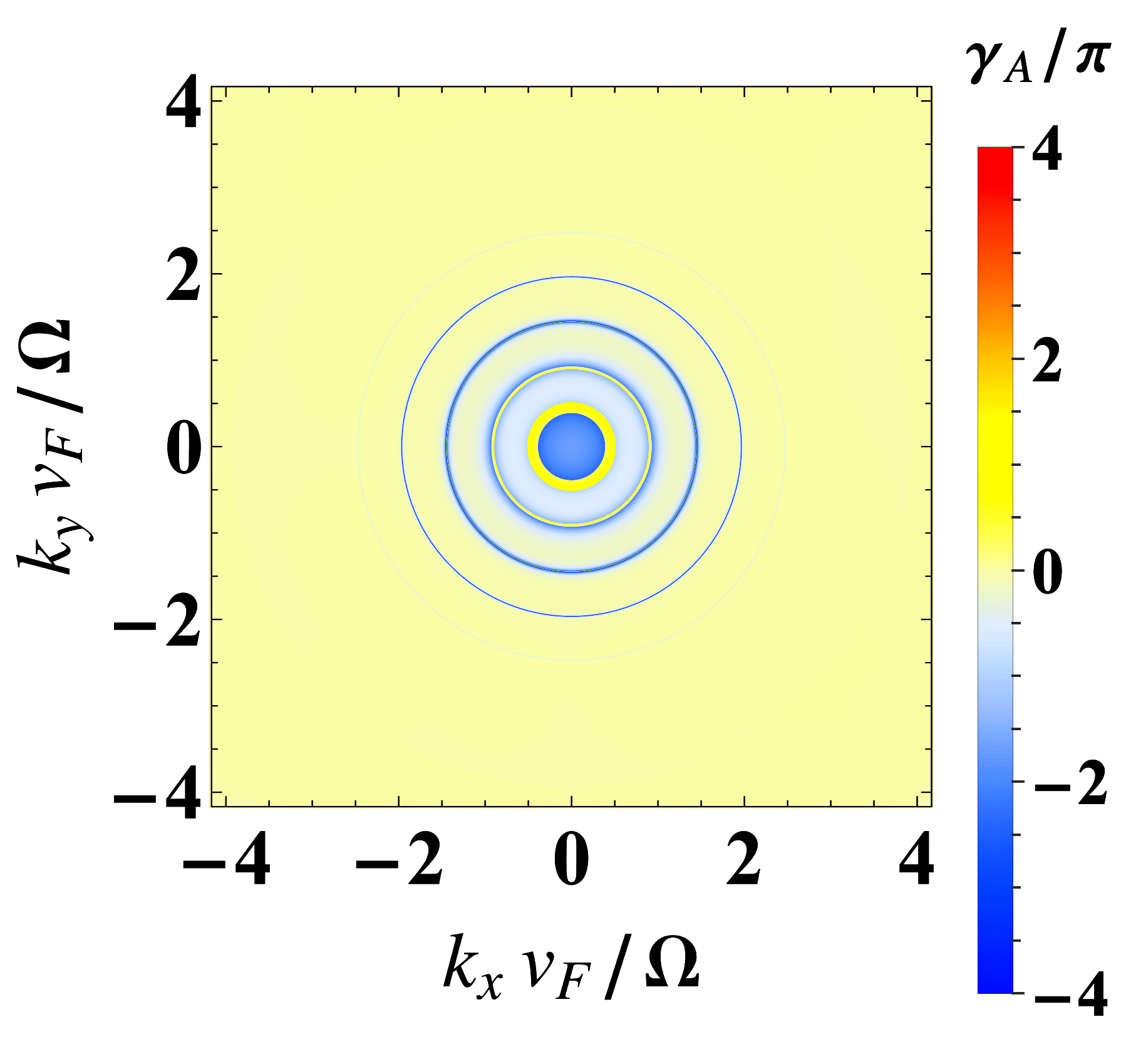}
\caption{Topological phases for circular polarized light in
the weak electric fields 
regime ($ev_FE_x/\hbar \Omega^2=ev_FE_y/\hbar \Omega^2=0.5$).
Panel  a)  presents the 
Berry phase as obtained from the Dirac-Bloch formalism.
Panel b) presents the Aharonov-Anandan phase obtained from solving
Eq. \eqref{eq:anandangeometric}. 
The Berry phase is non-trivial, $\gamma_{B,\boldsymbol{k}}=-\pi$, only within a 
circular light blue region centered at the Dirac point, with a radius determined by the 
magnitude of the electric field, while it remains trivial in the light yellow region. 
In contrast, the Aharonov-Anandan phase exhibits the same circular region but also 
features concentric rings, indicating multiphoton transitions.
}\label{fig:3}
\end{figure}

\section{Berry and Aharonov-Anandan phases in time-driven graphene} \label{sec:Numerical Methods}

To analyze this problem in detail, we consider the external driving of the graphene's effective Hamiltonian by an electromagnetic wave with angular frequency
$\Omega$, propagating along the $z$-direction, perpendicular to the
2D lattice plane and described by a vector potential
$\boldsymbol{A}(t)$ \cite{CARVALHO201724,Wright2009,Ishikawa2010,Ishikawa_2013,Lopez-Rodriguez21072010}.

Here we study four different cases: linearly polarized light
in the weak and strong field regimes and elliptically
polarized light in the weak and strong field regimes.
We assume that, for the cases in which light is
linearly polarized, the vector potential takes the form:
\begin{equation}
\boldsymbol{A}(t) = \left(\frac{E_x}{\Omega} \cos{(\Omega t)}, 0\right),
\label{eq:LineraPotential}
\end{equation}
where $E_x$ is the amplitude of the electric field.
In cases where the light excitation is elliptically polarized,
the vector potential is given by:
\begin{equation}
\boldsymbol{A}(t)
= \left( \frac{E_x}{\Omega} \cos{(\Omega t)},
\frac{E_y}{\Omega} \sin{(\Omega t)} \right),
\label{eq:vecpot}
\end{equation}
where $E_y$ represents the amplitude of the electric field
in the $y$-direction.

As explained in Appendix \ref{app:appendixA}, the time-dependent equation resulting from Eq. (\ref{eq:khamiltonian}) contains the dimensionless parameter,
\begin{equation}
    q_0=ev_F\sqrt{E_x^{2}+E_y^{2}}/\hbar \Omega^{2}
\end{equation}
that defines the limits between weak ($q_0<1$) and strong ($q_0>1$) external driving fields.    

{\color{black} For example, in Appendix \ref{app:appendixA} we show that for the linear polarized potential along the $x$ direction, the spinor components can be decoupled by using several unitary transformations. For one Dirac valley, the resulting wave function is,
\begin{equation}
  \mathbf{\Psi}_{\boldsymbol{k}}(t)=\exp\left[i\left(\frac{v_F k_y\phi}{\Omega}\sigma_0-\frac{\pi}{4} 
\sigma_y\right)\right]\begin{pmatrix}
      \chi_{1}(\phi) \\   \chi_{-1}(\phi) 
  \end{pmatrix} 
\end{equation}
where $\sigma_0$ is the $2\times 2$ identity matrix, $\phi=\Omega t/2$ and the  spinor $\mathbf{\chi}(\phi)=(\chi_{+1}(\phi),\chi_{-1}(\phi))^{\top}$ components  follow a Whittaker-Hill differential equation with complex coefficients, 
\begin{multline}
\chi_{\eta}''(\phi)+4\bigg[\left(\frac{v_F k_x}{\Omega}-q_0\cos(2\phi)\right)^{2}+\left(\frac{v_F k_y}{\Omega}\right)^{2}\\+i\eta q_0\sin(2\phi)\bigg] \chi_{\eta}(\phi)=0.
\label{eq:AlmostText}
\end{multline}
In the other Dirac valley, the components of the spinor are swapped.

Due to the complex parameter $i\eta q_0$, the previous equation behaves differently from the usual Whittaker-Hill equation. Its properties  have been scarcely studied even by mathematicians. Furthermore, even its limited, simplified version, the Mathieu equation with complex coefficients, remains an active area of mathematical research \cite{Kovacic,ZIENER2012,Malejki2021,Amore2021}.
This is because the complex part of the parameters
significantly affects the stability chart of the Mathieu equation \cite{Kovacic,ZIENER2012}.
Yet, several limiting cases provide useful insights on the shape of
the quasienergy spectrum and topological phases (see Appendix \ref{app:appendixA}). }

Finally, and to compare the two different topological phases,
we numerically solved Eq. (\ref{eq:mainH}) using the
two approaches: Dirac-Bloch and Floquet. Below we outline the
details of such calculations.

\subsection{Berry phase}
To calculate the Berry phase, we consider the Dirac- Bloch formalism 
\cite{VictorM2010,Kwan2020,Roy2013,Ishikawa_2013,DiMauro2022,
Ishikawa2010,Mikhailov_2008,Wright2009,Carvalho2018,CARVALHO201724,
Villari2018,tamashevich2022inhomogeneous,Li_2018}
using the following ansatz, 
\begin{equation}\label{eq:ansatz-wavefunction}
    \Psi_{\boldsymbol{k}}(t) = \sum_{\lambda} c^\lambda_{\boldsymbol{k}}(t)
\varphi^\lambda_{\boldsymbol{k}}(t)
\mathrm{e}^{-i \gamma^\lambda_{D,\boldsymbol{k}}(t)},
\end{equation}
The instantaneous eigenvalues and eigenvectors were found
from
\begin{equation}
\hbar v_F\boldsymbol{\sigma}
\cdot\left[\boldsymbol{k}+\frac{e}{\hbar}\boldsymbol{A}
(t)\right]\varphi^\lambda_{\boldsymbol{k}}
(t)=\epsilon^\lambda_{\boldsymbol{k}}
(t)\varphi^\lambda_{\boldsymbol{k}}(t).
\label{eq:khamilinsta}
\end{equation}
where we have used the Hamiltonian given by
Eq. (\ref{eq:khamiltonian}).

\begin{figure}
\centering
\large{a)}\includegraphics[width=0.9\linewidth]{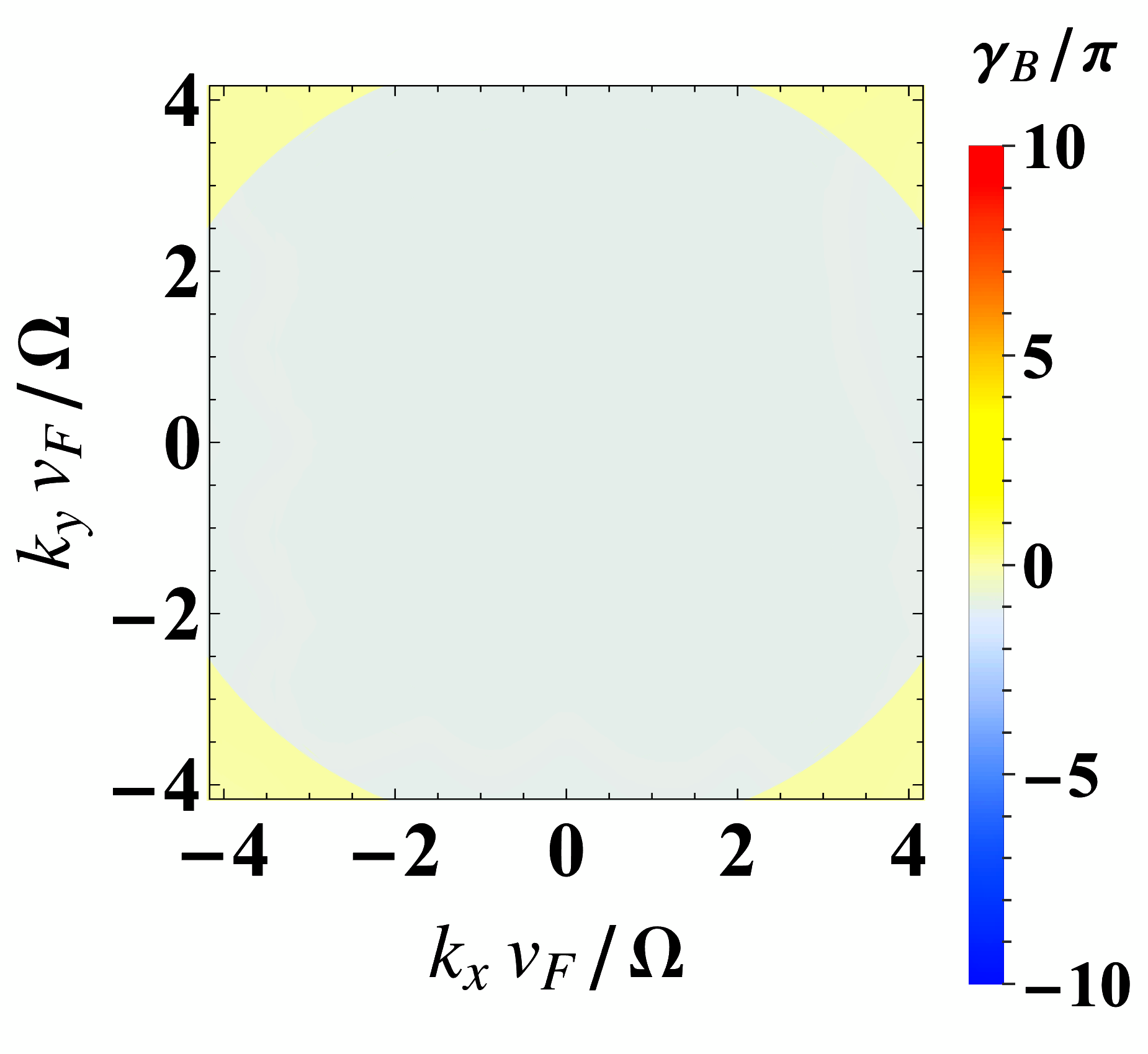}
\large{b)}\includegraphics[width=0.9\linewidth]{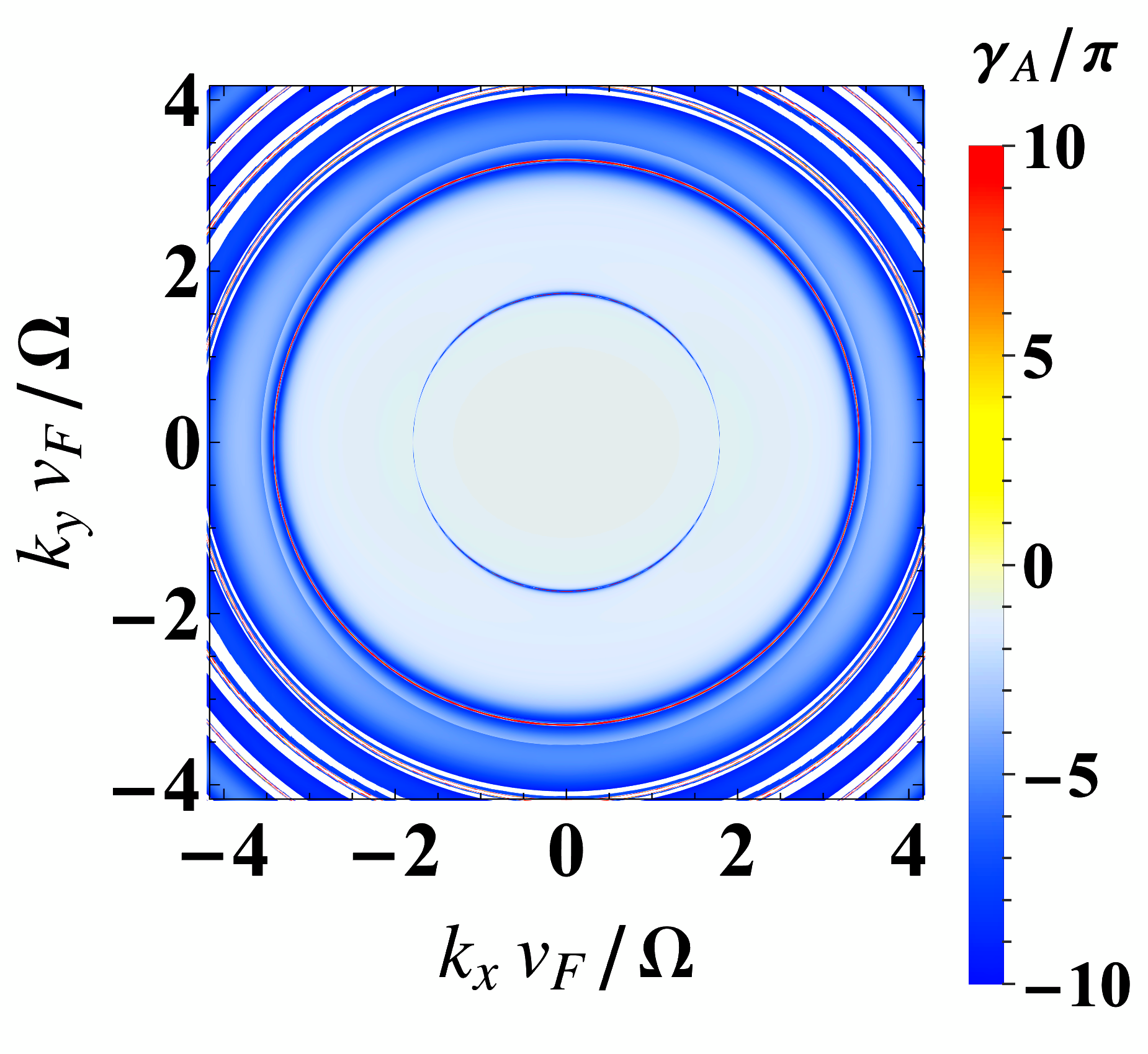}
\caption{Topological phases for circular polarized light in the strong 
electric fields regime ($ev_FE_x/\hbar \Omega^2=5.0$,
$ev_FE_y/\hbar \Omega^2=5.0$).
a) Berry phase,
b) Aharonov-Anandan phases.
As in Fig. \ref{fig:3}, the Berry phase is non-trivial
$\gamma_{B,\boldsymbol{k}}=-\pi$, only within a circular
light blue region centered at the Dirac point, with a radius
determined by the magnitude of the electric field.
The Aharonov-Anandan phase exhibits the same circular region
but also features concentric rings,
indicating multiphoton transitions.}\label{fig:4}
\end{figure}

The obtained eigenvalues and eigenvectors are then substituted 
into Eq. (\ref{eq:Geometrical-Berry-phase}) to compute the Berry 
phase. Figure \ref{fig:3} a) shows the resulting Berry phase in 
reciprocal space for the weak electric field regime using 
circular polarization, while Fig. \ref{fig:4} a) presents the 
results for the strong electric field regime. The cases with 
linear polarization are not shown, as they yield trivial Berry 
phases. A detailed discussion of these results can be found in Sec. \ref{sec:Discussion and conclusions}.

\begin{figure}
\centering
\includegraphics[width=0.9\linewidth]{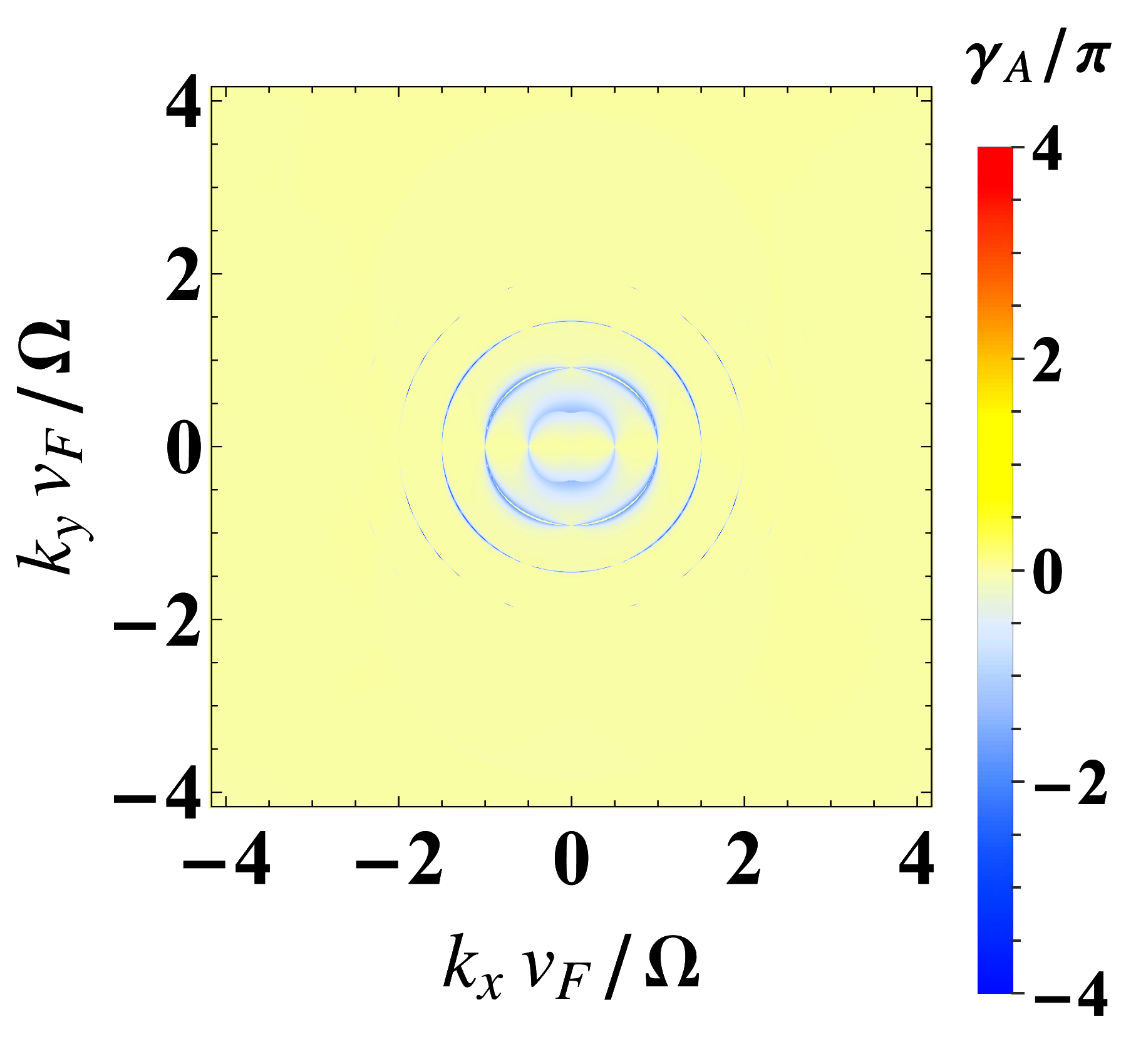}
\caption{Topological Aharonov-Anandan phases
for linear polarized light in the weak electric
fields regime ($ev_FE_x/\hbar \Omega^2=0.5$,
$ev_FE_y/\hbar \Omega^2=0.0$).
The phase vanishes for momenta in the light yellow region,
particularly for those parallel to the applied field
($k_y=0$). Note that for this case of linear polarization,
the Berry phase is always zero throughout the entire reciprocal
space; therefore, we do not include such a trivial plot.}
\label{fig:5}
\end{figure}

\begin{figure}
\centering
\includegraphics[width=0.9\linewidth]{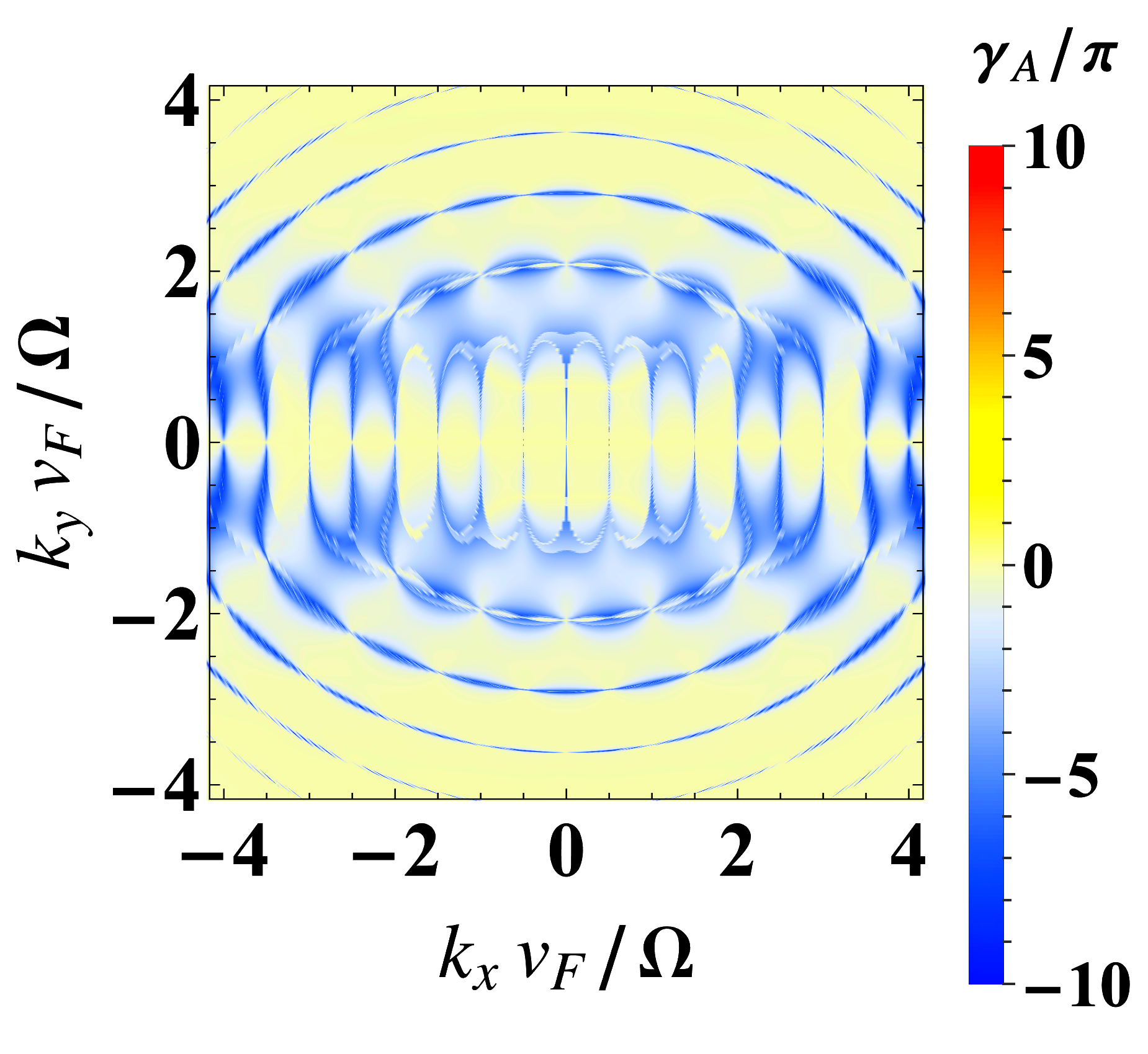}
\caption{Topological Aharonov-Anandan phases for linearly polarized light in the strong electric field regime
($q_0=ev_FE_x/\hbar \Omega^2=5.0$, $ev_FE_y/\hbar \Omega^2=0.0$).
As in the weak field regime, the phase vanishes for momenta parallel
to the applied field ($k_y=0$), the phase vanishes. Under these same conditions, the Berry phase is zero throughout the entire reciprocal
space and is therefore not shown.}
\label{fig:6}
\end{figure}
\subsection{Aharonov-Anandan phase}

To find the Aharonov-Anandan phase,
we numerically solve the Schr\"odinger equation
\eqref{eq:schroU} considering the Hamiltonian in Eq.
\eqref{eq:khamiltonian}.
The wave function was obtained using Floquet theory via
Eq. (\ref{eq:Floquet}), as detailed in Refs. 
\cite{PhysRevB.102.045134,Ibarra-Sierra_2022}.
The dynamical phase, $\gamma_{D,\boldsymbol{k}}$,
can be determined by first computing
$\mathcal{E}_{\boldsymbol{k}}^jT/\hbar$, 
which corresponds to the argument of the eigenvalues of
the evolution operator, given by
$U_{\boldsymbol{k}}(T)=\exp(-i/\hbar{H}_{e,\boldsymbol{k}} T)$.
Obtaining the corresponding eigenvectors, we use Eq. \eqref{eq:anandangeometric}
to calculate the Aharonov-Anandan phase. Finally, from
Eq. \eqref{eq:aharonovphases}, we derive the dynamical
phase, $\gamma_{\boldsymbol{k}}^j(T)$.

Figure \ref{fig:spectrumlowfield} displays the positive quasienergy,
$\mathcal{E}_{\boldsymbol{k}}^1T/\hbar$ ($j=1$), of the Floquet state
$\boldsymbol{\Psi}_{\boldsymbol{k}}^1(t)$ as a function of
$k_x$ and $k_y$ for (a) circularly and (b) linearly polarized
light in the weak-field regime. {\color{black} Notice that here the reciprocal space is scaled as $|\boldsymbol{k}|v_F/\Omega$ in order to use adimensional time units as suggestd by Eq. (\ref{eq:AlmostText}). }
Figure \ref{fig:spectrumlowfield} (a) and (b) show that in both cases the spectrum is distorted
compared to the free electron case from the Dirac point
outwards up to $v_F(k_x^2+k_y^2)^{1/2}/\Omega\approx 0.5$.
Beyond this
line the free-electron spectrum is recovered.
Under circularly polarized light, the quasi-energy spectrum
exhibits circular symmetry, whereas under linearly polarized
excitation, it is slightly stretched along the $k_x$ axis.
Additionally, the circularly polarized excitation induces a gap
at the Dirac point, while in the linear case, the Dirac point
remains gapless 

In Fig. \ref{fig:spectrumhighfield}
we observe the behavior of the positive quasienergy
$\mathcal{E}_{\boldsymbol{k}}^1T/\hbar$
as a function of $k_x$ and $k_y$ for
(a) circularly and (b) linearly polarized light in the strong field
regime.
In both cases the spectrum is distorted
in comparison with the free electron case from the Dirac point
outwards up to $v_F(k_x^2+k_y^2)^{1/2}/\Omega\approx 5.0$.
As in the low field regime, beyond this
line the free-electron spectrum is recovered.
The distortions are however much larger than in the
weak-field regime.
Under the circularly polarized light the
quasi spectrum newly has circular symmetry. The linearly polarized
excitation is strongly stretched along the $k_x$ axis and vertical
lines appear along $k_y$. Additionally, under
circularly polarized light the  gap that opens up
close to the Dirac point is much
larger than in the weak-field regime. In the linear
case the Dirac point remains.

The shape of the quasispectrum in Figs. \ref{fig:spectrumlowfield} and \ref{fig:spectrumhighfield} is determined
by the effective Whittaker and Hill diferential equation
with complex parameters, see Appendix \ref{app:appendixA}.

Finally, once the Floquet states are found using Eq. \eqref{eq:Floquet}, the 
Aharonov-Anandan phase is calculated by Eq. \eqref{eq:anandangeometric}.
Figs.  \ref{fig:3},\ref{fig:4},\ref{fig:5} and \ref{fig:6} present the resulting 
Aharonov-Anandan phases in reciprocal space for the weak and strong-field
regimes using different polarizations. The discussion of these results is given in Sec. \ref{sec:Discussion and conclusions}.

\section{Discussion and remarks} \label{sec:Discussion and conclusions}
Let us discuss the main features observed in the numerically obtained plots.  

\subsection{Weak external field}

In Figs. \ref{fig:3} we present the  Berry  and Aharonov-Anandan phases in
the weak-field regime using circular polarized light.
The Berry phase is different from zero, $\gamma_{B,\boldsymbol{k}}=-\pi$,
in a region centered at the tip of the Dirac cone, as seen in
Fig. \ref{fig:3} a) as a light blue circle. This justifies the
idea of assigning a topological charge to the Dirac cone \cite{Foa}, even if
the adiabatic condition is never met at the cone tip.
The  radius of the non-trivial topological Berry phase is determined by the
magnitude of the electric field, while it remains trivial in the light yellow 
region.
Meanwhile, as seen in Figs. \ref{fig:3} b), the Aharonov-Anandan phase is also 
different from zero in the same region but also presents several concentric rings. As we shall discuss later, such rings are
associated with transitions from the valence band to the conduction band.
The Aharonov-Anandan
phase in Fig. \ref{fig:4} b) exhibits a similar behavior.
However, in this case there are strong distortions within
the circle of radius $q_0=ev_FE/\hbar\Omega$. Outside this
circle the typical thin lines corresponding to the resonance
condition reappear as in the weak-field case.

In Fig. \ref{fig:5}, we present the case of linearly polarized light.
The Berry phase is not included, as it remains trivial for all values
of $\boldsymbol{k}$. Instead, the Aharonov-Anandan phase exhibits
a much richer and more intricate structure. For instance, along
the $k_x$-axis ($k_y=0$), the Aharonov-Anandan phase is zero,
a fact that will be explained later. Additionally, the rings observed
in Fig. \ref{fig:3} are now deformed,
transforming into ellipses that eventually recover their
circular shape.
 
\subsection{Strong external field}

Fig. \ref{fig:4} presents the Berry and Aharonov-Anandan phases for the strong external 
field case with circular polarization. Its structure is similar to the case of a 
weak-field except that the radius is expanded due to the bigger magnitude of the 
electric field. Again the Berry phase is non-trivial only within a circle around 
the Dirac cone. In contrast, the Aharonov-Anandan phase seen in  Fig. \ref{fig:4} 
b) exhibits concentric rings, indicating multiphoton transitions. 

The Aharonov-Anandan phases shown in Fig. \ref{fig:6} exhibit a rather
intriguing behavior.
First the phase is zero along the $k_y=0$, meaning that for
states with momentum only in the direction of the applied field,
the phase remains zero.
For states with $k_y \neq 0$, the rings observed in Fig. \ref{fig:4}b) transform 
into ellipses with a modulation that creates a pearl-necklace-like structure.
In this case, the Berry phase does not provide information about these structures, as it remains zero throughout the reciprocal space.

\subsection{Remarks on the phases}

\begin{figure*}[t]
    \centering
    \large{a)}\includegraphics[width=0.35\linewidth]{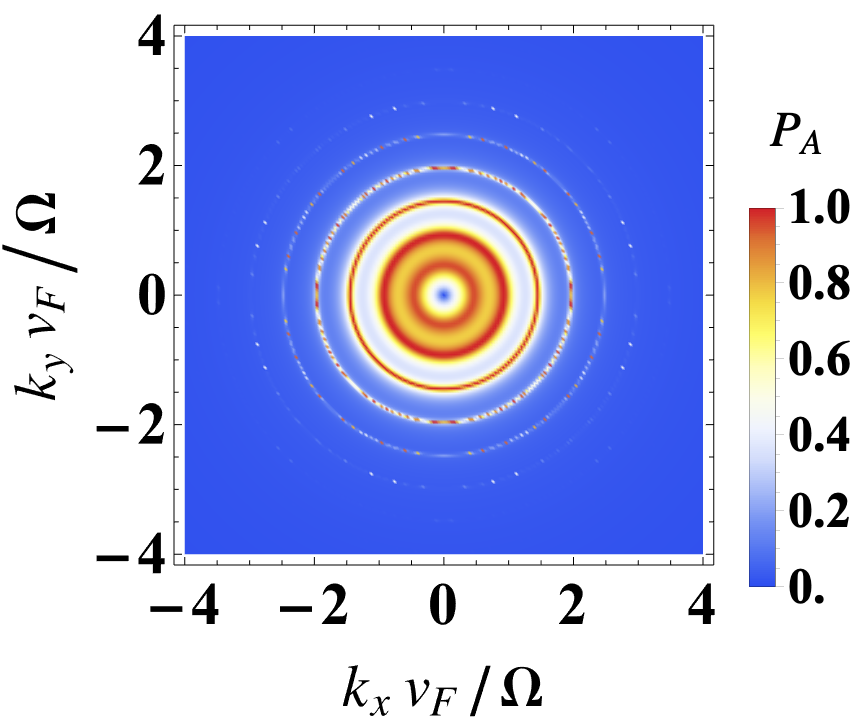}
    \large{b)}\includegraphics[width=0.35\linewidth]{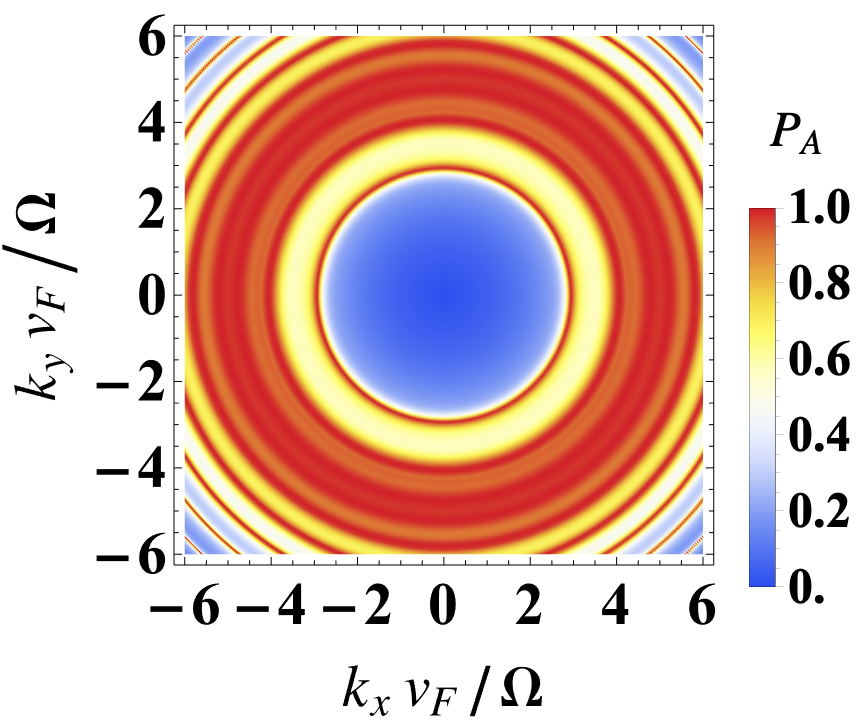}
    \large{c)}\includegraphics[width=0.35\linewidth]{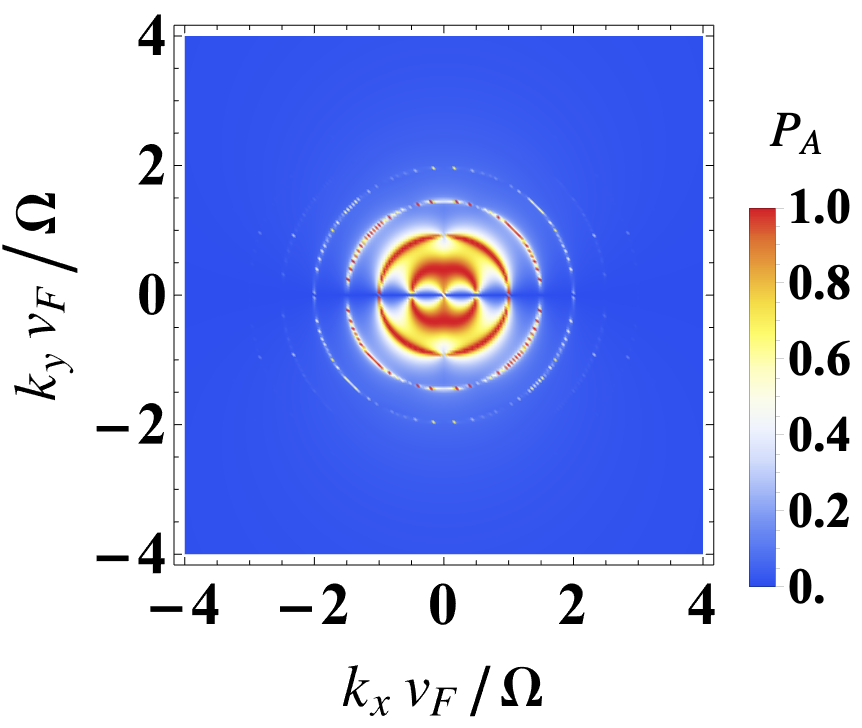}
    \large{d)}\includegraphics[width=0.35\linewidth]{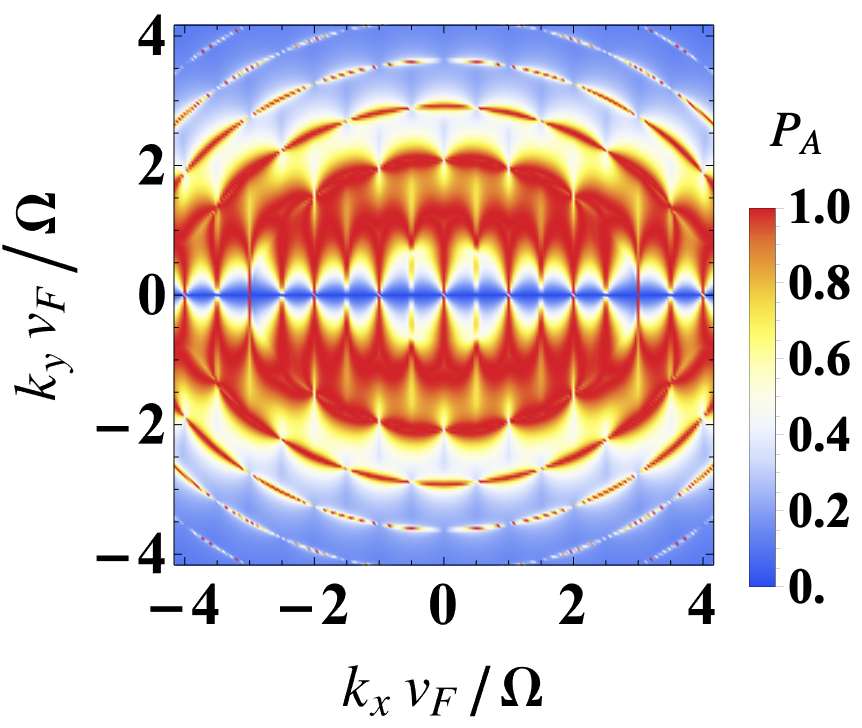}
    \caption{Transition probability as a function of
    $k_xv_f/\Omega$ and $k_yv_f/\Omega$ for
    (a) circularly
    polarized light in the weak electric field regime
    ($q_0=ev_FE_x/\hbar \Omega^2=0.5$,
    $ev_FE_y/\hbar \Omega^2=0.5$) and
    (b) circularly
    polarized light in the strong electric field regime
    ($q_0=ev_FE_x/\hbar \Omega^2=5.0$,
    $ev_FE_y/\hbar \Omega^2=5.0$),
    (c) linearly
    polarized light in the weak electric field regime
    ($q_0=ev_FE_x/\hbar \Omega^2=0.5$,
    $ev_FE_y/\hbar \Omega^2=0.0$),
    (d) linearly
    polarized light in the strong electric field regime
    ($q_0=ev_FE_x/\hbar \Omega^2=5.0$,
    $ev_FE_y/\hbar \Omega^2=0.0$).}
    \label{fig:7}
\end{figure*}

In summary, for linearly polarized light, the Berry phase yields trivial results.
In contrast, for circularly polarized light, the Berry phase is nonzero only in a 
region around the Dirac cone where the adiabatic condition is not met. This region, 
determined in reciprocal space by the strength and frequency of the external drive, 
also exhibits a nonzero Aharonov-Anandan phase. However, the Aharonov-Anandan phase 
reveals much more structure, which can be understood by considering band 
transitions.

It can be shown from Eq. \eqref{eq:instantstate}
and the Stokes theorem
that the Berry phase in Eq. 
\eqref{eq:Berry_phase_definition}
can be rewritten as:
\begin{multline}
\gamma_{B,\boldsymbol{k}}^{\lambda}
=\frac{1}{2}\oint_C
\frac{K_y(t)dK_x(t)-K_x(t)dK_y(t)}{K_x^2(t)+K_y^2(t)} \\
=-\frac{1}{2}\int_S dK_x dK_y \bigg[
\frac{\partial}{\partial K_x}\left(\frac{K_x}
{K_x^2+K_y^2}\right) \\
+\frac{\partial}{\partial K_y}\left(\frac{K_y}
{K_x^2+K_y^2}\right)
\bigg],\label{eq:berryphasestokes}
\end{multline}
where $K_x(t)$ and $K_y(t)$ are the components
of $\boldsymbol{K}(t)$.
Here, $C$ is the contour traced by the vector potential
over one period $T$, and $S$ is the surface enclosed
by this contour.

This integral evaluates to $-\pi$ as long as the
singularity at $\boldsymbol{k}=\boldsymbol{0}$
lies inside the contour $C$.
Conversely, if the singularity lies outside the contour
$C$, the right-hand side
of Eq. \eqref{eq:berryphasestokes} vanishes trivially. 
In the specific case where the vector potential
is given by  Eq. \eqref{eq:vecpot},
the Berry phase is $-\pi$ if the 
ellipse centered at $(k_x,k_y)$, with semi-minor and
semi-major axes equal to $eE_x/\hbar\Omega$
and $eE_y/\hbar\Omega$, respectively,
contains the origin in reciprocal space.
Otherwise, the Berry phase vanishes.
In this way, when graphene is subject to
linearly polarized light ($E_x\ne 0$ and $E_y=0$)
the eccentricity
of the ellipse is zero and the Berry phase
vanishes all over the reciprocal space.
Instead, under circularly polarized light
the eccentricity of the ellipse
is one.
This is confirmed in the numerical simulations; we observe in
Figs. \ref{fig:3} a) and \ref{fig:4} a) that
the Berry phase is $-\pi$ inside
the circle of radius
$eE_x/\hbar\Omega=eE_y/\hbar\Omega$
and $0$ elsewhere.

Regarding the Aharonov-Anandan phase we note the following points:
\begin{enumerate}
\item The rings that form in the Aharonov-Anandan phase
in the weak field regime under circularly polarized
excitation shown in Fig. \ref{fig:3}
are consistent with the resonance condition
$2\epsilon_{\boldsymbol{k}}=2q\hbar \omega$
for $q=1,2,\dots$ photons as it was shown
in the rotating wave approximation \cite{PhysRevB.102.045134}.

\item The Aharonov-Anandan phase also presents
rings in the weak field regime
but under linearly polarized light,
as can be seen in Fig. \ref{fig:5}.
However, in this case the rings disappear
near the region where $k_y=0$ where
the transitions between the valence and
conduction bands are forbidden as we demonstrate
below.
When $k_y=0$ and under linearly polarized light ($E_y=0$)
the evolution operator is obtained from
the Hamiltonian in Eq. \eqref{eq:khamiltonian}
giving
\begin{equation}
U(t)=\exp\left[-iv_F
\alpha(t)
\sigma_x
\right],
\label{eq:kxsol}
\end{equation}
where
\begin{equation}
\alpha(t)=k_x t+\frac{eE_x}{\hbar\Omega^2}\sin(\Omega t).
\end{equation}
Under this very particular set of conditions,
the conduction and valence band states
\begin{equation}
\psi_C = \frac{1}{\sqrt{2}}\left(
\begin{array}{c}
1\\ 1
\end{array}
\right), \,\,\,\,\,\,
\psi_V = \frac{1}{\sqrt{2}}\left(
\begin{array}{c}
1\\ -1
\end{array}
\right),
\end{equation}
also correspond to the Floquet states
with quasi energies $\mathcal{E}^{\pm 1}=\pm \hbar v_F k_x$.
The transition probabilities between these
two states vanish, namely
\begin{equation}
\left\langle \psi_C\left\vert
U(t)\right\vert\psi_V\right\rangle=0.
\end{equation}
Moreover, assuming that the system
begins in the conduction band, the time-dependent wave function
is given by
\begin{equation}
\psi(t) =\frac{1}{\sqrt{2}}\left(
\begin{array}{c}
\mathrm{e}^{i\alpha(t)}\\ -\mathrm{e}^{i\alpha(t)}
\end{array}
\right)
\end{equation}
The Andandan phase of Eq. \eqref{eq:anandangeometric} in this case
vanishes because $w_{\boldsymbol{k}}^j=1$ for any of the bands
and therefore $dw_{\boldsymbol{k}}^j=0$. This can be clearly
seen in Figs. \ref{fig:5} and \ref{fig:6} where $\gamma_{A,\boldsymbol{k}}^j=0$
in the $k_x$-direction where $k_y=0$.
A similar argument can be made in the
bipartite lattice base
\begin{equation}
\psi_A = \left(
\begin{array}{c}
1\\ 0
\end{array}
\right), \,\,\,\,\,\,
\psi_B = \left(
\begin{array}{c}
0\\ 1
\end{array}
\right),
\end{equation}
where the Aharonov-Anandan phase is given
by $\gamma_{A,\boldsymbol{k}}=-\int_0^T\tan{\alpha(t)}(d\alpha(t)/dt) dt=0$.

{\color{black}An alternative approach is shown in Appendix \ref{app:appendixA} as it is possible to
derive the trivial nature of the phases for $k_y=0$ given by Eq. (\ref{eq:kxsol}) by integrating
the Whittaker-Hill Eq. (\ref{eq:AlmostText}).}

\item This also explains some other features
of the Aharonov-Anandan phase
in the strong field regime
observed in Fig. \ref{fig:6} as
the vertical lines with values of
$\gamma_{A,\boldsymbol{k}}\ne 0$
in the vicinity of $-\pi$
in the region of reciprocal space where
the field dominates ($k< eE_x/\hbar \Omega$).
In this region the symmetry in reciprocal space
is strongly broken by the field producing
vertical lines. As we move far from this
region ($k> eE_x/\hbar \Omega$) the
symmetry is restored and the transition lines
that comply with the resonance condition
$2\epsilon_{\boldsymbol{k}}=2q\hbar \omega$
recover the circular symmetry. However, the distortions
of the oscillating electric
field along the $k_x$-direction are still observed
as small modulations in the transition lines.

\item Thus in general, it can be established
that $\gamma_{A, \boldsymbol{k}}\ne 0$ ($\approx -\pi$)
where the resonance condition is met
($2\epsilon(\boldsymbol{k})=q\hbar \omega $ where $q=1,2\dots , n$) and $\gamma_{A, \boldsymbol{k}}= 0$ between
the transition lines.
It is then natural to ask ourselves why is there
a connection between the transitions and the
Aharonov-Anandan phase.
This question can be answered through
the following argument.
Weather in the conduction-valence band basis
$\psi_V$, $\psi_C$ or the bipartite lattice basis
$\psi_A$, $\psi_B$
the Floquet wave function must be of the form
\begin{equation}
\psi^j_{\boldsymbol{k}}(t)
= \exp\left(-i\frac{\mathcal{E}^j_{\boldsymbol{k}}}{\hbar}t\right)
\mathrm{e}^{i\chi (t)}\left(
\begin{array}{c}
\mathrm{e}^{i\beta(t)} a(t)\\
b(t)
\end{array}
\right)
\end{equation}
where, in accordance with the Floquet theorem,
$\mathcal{E}^j_{\boldsymbol{k}}$ is the quasi energy,
and $\chi(t)$, $\beta(t)$, $a(t)$ and $b(t)$ real
numbers and are periodic
with period $T$.
The Aharonov-Anandan phase that arises from this wave function
is
\begin{equation}
\gamma_{A,\boldsymbol{k}}=\frac{1}{2}
\int_0^T\dot{\beta}(t)\frac{a^2(t)}{a^2(t)+b^2(t)}dt.
\end{equation}
Far away from the resonant condition
$a(t)$ and $b(t)$ are constants. In this case
the Aharonov-Anandan phase is
\begin{equation}
\gamma_{A,\boldsymbol{k}}=\frac{1}{2}
\frac{a^2(t)}{a^2(t)+b^2(t)}
\int_0^T\dot{\beta}(t)dt=0.
\label{eq:anandan-no-transtion}
\end{equation}
If, on the contrary, the resonant condition is met, $a(t)$ and $b(t)$ are time-dependent and the behaviour of the Anandan phase is non-trivial.

In Fig. \ref{fig:7}, we have used the difference between the 
maximum and minimum probabilities of the Floquet state being in 
site $A$ ($\psi_A$) during a period $T$ as a measure of the 
transition probability.
Similar results are obtained by subtracting the minimum
and maximum probabilities of being in the conduction band, 
though they are not shown here. Given the strong resemblance 
between these density plots and the Aharonov-Anandan phase, 
along with the result of Eq. \eqref{eq:anandan-no-transtion}, it 
is evident that $\gamma_{A,\boldsymbol{k}}$ encodes information 
about the transitions induced by the periodic drive.

\item The connection between the Berry and Aharonov-Anandan phases can also be worked out from a perturbative approach of the time-evolution operator as shown in Appendix \ref{Appendix: Approximation-Time-Evolution}. 

\end{enumerate}

{\color{black}
\section{Relationship between currents
and the Berry and Aharonov-Anandan phases}\label{Sec:Currents}

In this section we demonstrate that the interaction between electrons in the graphene lattice and the electromagnetic field gives rise to vortices associated with the Berry and Aharonov-Anandan phases. The structure of these phases depends strongly on the field regime (weak or strong). Furthermore, we establish a connection between these currents and the corresponding geometric phases.

We can calculate the currents associated with the instantaneous
eigenstates of Eq. \eqref{eq:insteigstates}
using the components of the velocity operators
$v_F\sigma_x$ and $v_F\sigma_y$ as
\begin{eqnarray}
J_{x,\boldsymbol{k},\xi}^\lambda &=& -nev_F\varphi_{\boldsymbol{k},\xi}^{\lambda\dagger}
\sigma_x\varphi_{\boldsymbol{k},\xi}^{\lambda}
=  -nev_F\lambda \frac{K_x(t)}{K(t)},\\
J_{y,\boldsymbol{k},\xi}^\lambda &=& -nev_F\varphi_{\boldsymbol{k},\xi}^{\lambda\dagger}
\sigma_y\varphi_{\boldsymbol{k},\xi}^{\lambda}
=  -nev_F\lambda \frac{K_y(t)}{K(t)},
\end{eqnarray}
where $n$ represents the electron surface density.
Thus, the Berry phase can be expressed in terms of the
current components as
\begin{multline}
\gamma_{B,\boldsymbol{k},\xi}^\lambda
=-\frac{\xi}{\left(nev_F\right)^2}\oint_B
\left(J_{x,\boldsymbol{k},\xi}^\lambda\,\, dJ_{y,\boldsymbol{k},\xi}^\lambda
-J_{y,\boldsymbol{k},\xi}^\lambda\,\, dJ_{x,\boldsymbol{k},\xi}^\lambda
\right)\\
=-\xi \int d\theta_{\boldsymbol{k}}(t)=-\xi \Delta\theta_{\boldsymbol{k}}
\end{multline}
where $\theta_{\boldsymbol{k}}(t)$ is given 
in Eq. \eqref{eq:berryangle}.
The preceding equation shows that the Berry phase is,
in fact, the vorticity of the instantaneous density
current. As we have seen above, it is entirely defined
by the vector potential and therefore exhibits rather
trivial behavior.

Conversely, the electrical currents associated
with the Floquet states can be calculated as
\begin{eqnarray} 
J_{x,\boldsymbol{k}}^j(t) = -n e v_F
\boldsymbol{\Psi}_{\boldsymbol{k}}^{j\dagger}(t)\sigma_x
\boldsymbol{\Psi}_{\boldsymbol{k}}^{j}(t),
\label{eq:currents_componentsx}
\\
J_{y,\boldsymbol{k}}^j(t) = -n e v_F
\boldsymbol{\Psi}_{\boldsymbol{k}}^{j\dagger}(t)\sigma_y
\boldsymbol{\Psi}_{\boldsymbol{k}}^{j}(t).
\label{eq:currents_componentsy}
\end{eqnarray}
These currents are depicted in Fig. \ref{fig:8}, where several snapshots of the current vector field generated from Eqns. \eqref{eq:currents_componentsx} and \eqref{eq:currents_componentsy} are presented for different times. These are superimposed on the density plot of the quasienergy spectrum in reciprocal space, showing a clear relationship. In Fig. \ref{fig:8}, the average current after one cycle is also presented, showing, for example, that the behavior in the $x$-direction is very different from that in other directions.
To understand the connection between the phases and currents, we observe that Eqns.
\eqref{eq:currents_componentsx} and 
\eqref{eq:currents_componentsy}
can be rewritten as,
\begin{eqnarray} \label{eq:polar}
J_{x,\boldsymbol{k}}^j(t) = -2 n e v_F
\left| \Psi_{\boldsymbol{k},\xi,2}^j(t)\right|
\left|\Psi_{\boldsymbol{k},\xi,1}^j(t) \right|
\cos  [ \Theta_{\boldsymbol{k}}(t)],\\
J_{y,\boldsymbol{k}}^j(t) = -2\xi n e v_F
\left|\Psi_{\boldsymbol{k},\xi,2}^j(t)\right|
\left|\Psi_{\boldsymbol{k},\xi,1}^j(t)\right|
\sin  [ \Theta_{\boldsymbol{k}}(t)],
\end{eqnarray} 
from where,
\begin{equation}
    \frac{J_{y,\boldsymbol{k}}^j(t)}{J_{x,\boldsymbol{k}}^j(t)}=\xi \tan[  \Theta_{\boldsymbol{k}}(t)]
\end{equation}
Next we integrate over a period $T$ the previous expression to obtain,
\begin{equation}
    \Delta\Theta_{k}= \xi \int_{0}^{T} \frac{d}{dt}\Theta_{\boldsymbol{k}}dt= \xi \int_{0}^{T}\frac{d}{dt}\left\{\tan^{-1}  \left( \frac{J_{y,\boldsymbol{k}}^j(t)}{J_{x,\boldsymbol{k}}^j(t)}\right) \right\} dt
    \label{eq:currentangle}
\end{equation}
{\color{black}
proving a relationship between $\Delta\Theta_{\boldsymbol{k}}$,
the relative phase between spinor components after one
driving cycle, and the ratio of the current components.

Eq. \eqref{eq:currentangle} suggests that a current vortex can
produce a non-trivial Aharonov-Anandan phase and vice versa,
i.e., a photon-induced
transition can produce a current vortex. We can prove this idea more
formally by using the expression for the currents given by Eqs.
\eqref{eq:currents_componentsx} and \eqref{eq:currents_componentsy}.
Together with Eqs. \eqref{eq:wdef} and \eqref{eq:anandangeometric},
the Aharonov-Anandan phase can be expressed in terms of the
current components as
\begin{equation} 
\gamma_{A,\boldsymbol{k}}^j=-\frac{1}{(2nev_F)^2}
\oint_C\frac{J_{x,\boldsymbol{k}}^j\,\,dJ_{y,\boldsymbol{k}}^j
-J_{y,\boldsymbol{k}}^j\,\,d J_{x,\boldsymbol{k}}^j}
{\left\vert\Psi_{\boldsymbol{k},1}^j(t)\right\vert^2}.
\end{equation}
This equation states that the Aharonov-Anandan phase is also
connected to the vorticity of the electrical currents. However,
these currents, being derived from the Floquet states, exhibit
much greater complexity than the instantaneous ones, as can be
readily observed in Fig. \ref{fig:8}. Notably, the current forms
vortices within the region of dominant field strength
($k<eEx/\hbar \Omega$).
In contrast, outside this region ($k>eEx/\hbar \Omega$),
where the free carrier energy prevails over the field,
the currents exhibit a more uniform behavior.
Therefore, the notion that the emission and absorption 
of photons generate vorticity in the currents is in
accordance with the results of the previous sections.

We note that the integrand numerator can be written as a wedge product ($\wedge$),
\begin{equation} \label{eq:anandan-wedge}
\gamma_{A,\boldsymbol{k}}^j=- \frac{1}{(2nev_{F})^{2}} \oint_{C} \frac{[\boldsymbol{J}_{\boldsymbol{k}}^{j} \wedge d \boldsymbol{J}_{\boldsymbol{k}}^{j}]_{z
}}{\left\vert\Psi_{\boldsymbol{k},1}^j(t)\right\vert^2}.
\end{equation}
Using Eqs. \eqref{eq:polar} it can be written in terms of the population oscillations and phase difference,
\begin{multline}
\gamma_{A,\boldsymbol{k}}^{j}
=- \frac{1}{(2nev_{F})^{2}} \oint_{C} \frac{J_{x,\boldsymbol{k}}^j\,\,dJ_{y,\boldsymbol{k}}^j
-J_{y,\boldsymbol{k}}^j\,\,d J_{x,\boldsymbol{k}}^j}{J_{x, \boldsymbol{k}}^{j}(t)^{2}} \frac{J_{x, \boldsymbol{k}}^{j}(t)^{2}}
{\left\vert\Psi_{\boldsymbol{k},1}^j(t)\right\vert^2}\\
= - \frac{1}{(2env_{F})^{2}} \oint_{C} \frac{J_{x, \boldsymbol{k}}^{j}(t)^{2}}
{\left\vert\Psi_{\boldsymbol{k},1}^j(t)\right\vert^2}  d \left( \tan^{-1} \left( \frac{J_{y, \boldsymbol{k}}^{j}(t)}{J_{x, \boldsymbol{k}}^{j}(t)} \right) \right)\\
= - \xi \oint_{C} \left\vert\Psi_{\boldsymbol{k},2}^j(t)\right\vert^2 \cos^{2}(2  \Theta_{\boldsymbol{k}}(t)) d \Theta_{\boldsymbol{k}}(t).
\end{multline}

The previous equation confirms the relationship between transitions and the Aharonov-Anandan phase.  On the other hand, we can define an instantaneous vorticity term $\boldsymbol{\bar{\omega}}(t)$ in the reciprocal space, as 

\begin{equation} \label{eq:vortex}
    \begin{split}
  \boldsymbol{\bar{\omega}}(t) dt&\equiv    \nabla_{\boldsymbol{K}(t)} \times \boldsymbol{J}_{\boldsymbol{k}}^{j}(t) dt\\
&= \epsilon_{lmn} \partial_{K_{m}(t)} J_{n,\boldsymbol{k}}^{j}(t) dt \\
&=\epsilon_{lmn} \left(\frac{d{K_{m}(t)}}{dt}\right)^{-1}\frac{d J_{n,\boldsymbol{k}}^{j}(t)}{dt} dt \\
&= (\dot{K}^{-1}_{x}(t),\dot{K}^{-1}_{y}(t)) \wedge d \boldsymbol{J}_{\boldsymbol{k}}^{j}  
    \end{split}
\end{equation}
where $\epsilon_{lmn}$ is the Levi-Civita tensor. A comparison between Eq.\eqref{eq:vortex} and Eq.\eqref{eq:anandan-wedge} reveals that whenever $\gamma_{A, \boldsymbol{k}}^{j} \neq 0$, it necessarily implies $dJ_{\boldsymbol{k}}^{j} \neq 0$, and thus the vorticity is nonvanishing. Physically, this suggests that the interaction between photons and electrons induces currents, which in turn generate magnetic fields responsible for the emergence of vortices.

\begin{figure*}[t]
    \centering
    \large{a)}\includegraphics[width=0.4\linewidth]{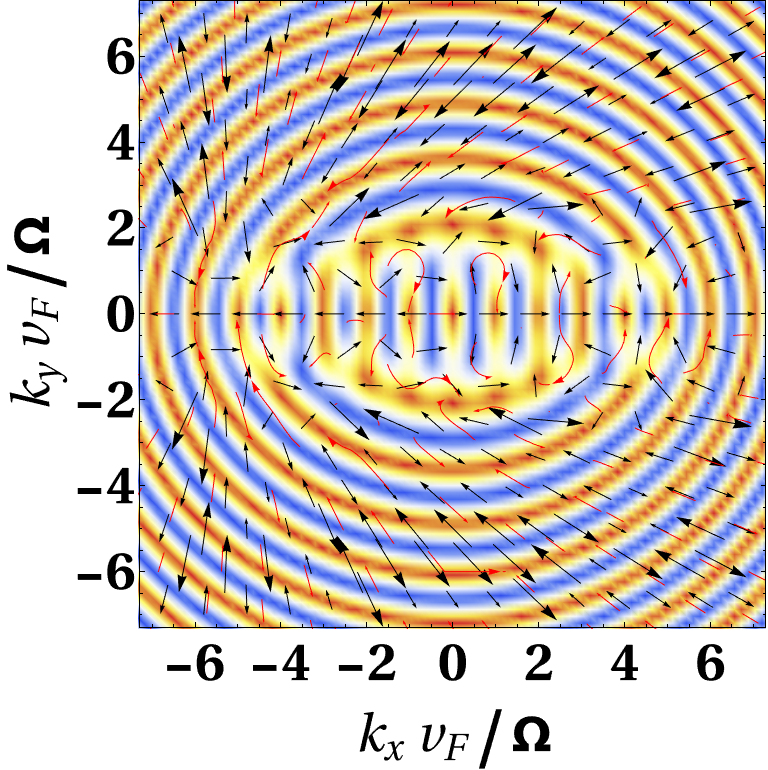}
    \large{b)}\includegraphics[width=0.4\linewidth]{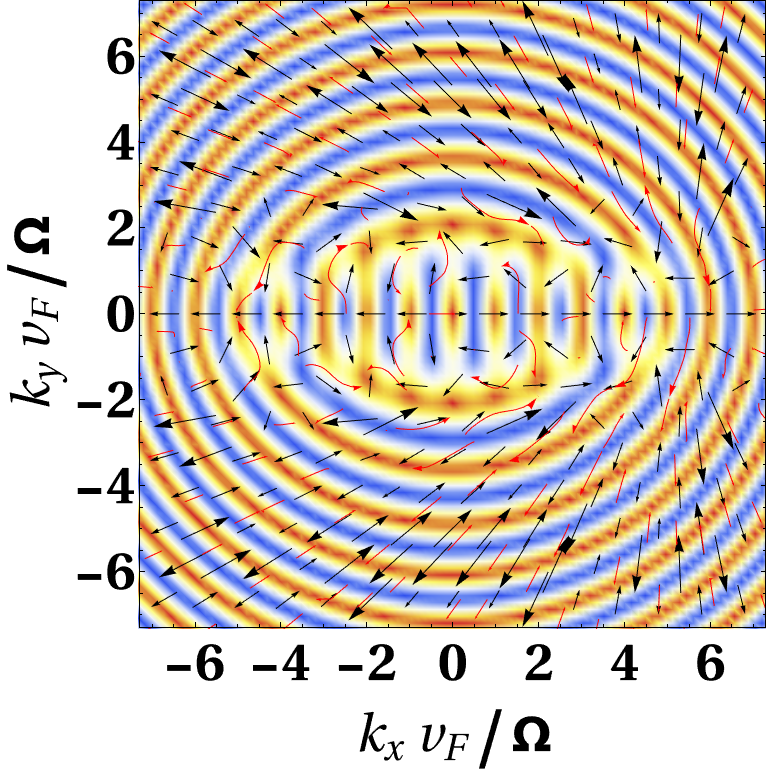}\\
    \large{c)}\includegraphics[width=0.4\linewidth]{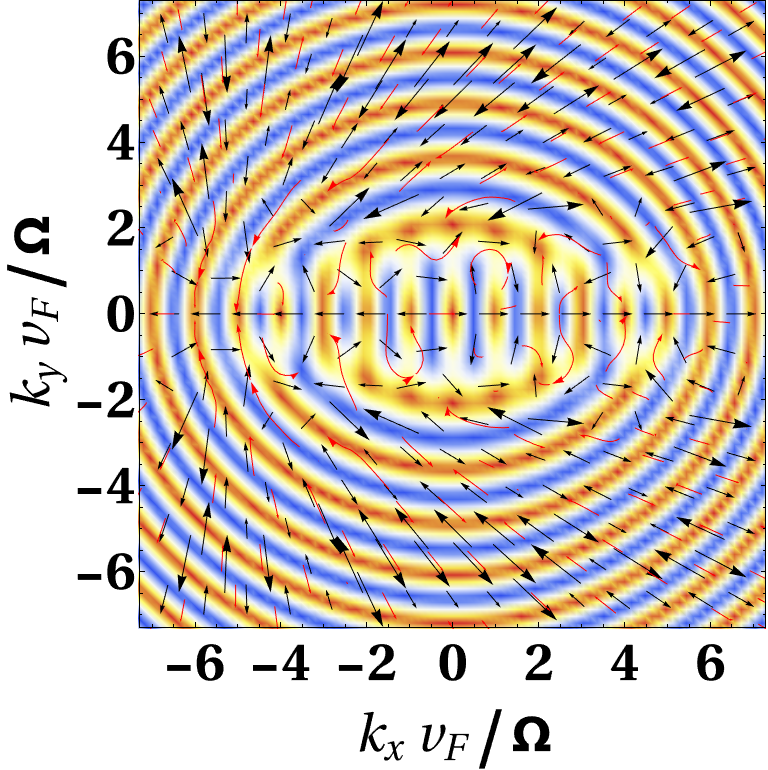}
    \large{d)}\includegraphics[width=0.4\linewidth]{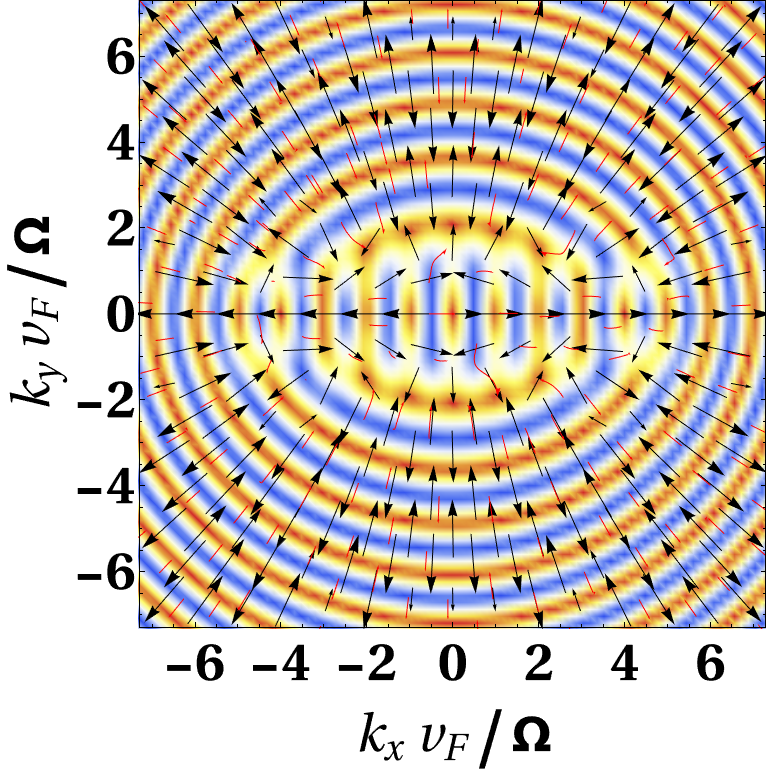}
    \caption{\textcolor{black}{
The current vector field (black arrows) $\boldsymbol{J}_{\boldsymbol{k}}{(t)}/(-nev_F)$ given by Eqns. \eqref{eq:currents_componentsx} and \eqref{eq:currents_componentsy} is shown, along with streamlines (red curves) as a visual aid, for different reference times: (a) $t=0$, (b) $t=T/2$, (c) $t=T$, and (d) the time-averaged current over one period, i.e., $\langle \boldsymbol{J}_{\boldsymbol{k}}{(t)} \rangle_T/(-nev_F)$.  In the background, the Floquet quasienergy spectrum is displayed. The system is considered in the strong electric field and linearly polarized regime, with parameters $q_0=ev_F E_x / \hbar \Omega^2=5.0$ and $ev_F E_y / \hbar \Omega^2=0.0$.}}
    \label{fig:8}
\end{figure*}
}

}

\section{Conclusions} \label{sec:Conclusions}

Although many solid-state systems are gapless, the Berry topological phase is derived from a trajectory in the reciprocal space without accounting for the breakdown of the adiabatic approximation, which requires a gap. {\color{black} While the Berry phase can always be computed using well-known formulas in reciprocal space, it does not offer insight into the limits of adiabaticity.} Determining these limits requires a time-dependent approach, as demonstrated here for graphene under electromagnetic radiation. We also show that the Aharonov-Anandan phase serves as a valuable alternative for weak fields, as it coincides with the Berry phase while offering additional insights, such as identifying regions of multiphoton transitions. 

The Berry and dynamical phases derived from the Dirac-Bloch formalism are solely determined by the contour of the vector potential. Consequently, these topological phases can be engineered by simply adjusting the polarization state and overall shape of the incident light. In contrast, tailoring the Aharonov-Anandan phase and its associated dynamical phases, which emerge from the Floquet formalism, is more challenging, as their evolution is entirely governed by the time-dependent dynamics of the wave function.

In stark contrast with the Berry phase, that represents
the phase acquired by the wave function remaining in the same sate, the Aharonov-Anandan phase encodes information
of the transition between levels. {\color{black} It was also proven that a non-trivial Aharonov–Anandan phase implies the presence of currents with vortices. Numerical simulations performed in graphene under electromagnetic radiation support this result.}

{\color{black} In conclusion, we have related non-adiabatic topological phases with electronic transitions and currents induced by the applied electromagnetic field in gapless solid-state systems.}

\acknowledgments
AK is indebted to IFUNAM for their hospitality
and was financially supported by Departamento de Ciencias
Básicas UAM-A Grant No. 2232218. AJEC and GGN thanks the CONAHCyT fellowship (No. CVU 1007044) and the Universidad Nacional Autónoma de México (UNAM) for providing financial support (UNAM DGAPA PAPIIT IN101924 and CONAHCyT project 1564464). The authors acknowledge and express gratitude to Carlos Ernesto L\'opez Natar\'en from Secretaria T\'ecnica de C\'omputo y Telecomunicaciones for his valuable support to implement high-performance numerical calculations. We also thank LSCSC-LANMAC for providing access to their HPC server, where part of the computational simulations for this work were performed.

\appendix

\section{Quasienergy spectrum of graphene under time-dependent radiation.\label{app:appendixA}}

In this section, we study the time-dependent Eq. \eqref{eq:scrhodinger1}, focusing on the most interesting result observed in the numerical simulations: the case of linear polarization. The evolution of the wave function is given by solving Eq. \eqref{eq:scrhodinger1}. The solution can be found by using the Floquet theory. This requires to uncouple the spinor components in the Dirac time-dependent equation. Consider the linear polarized case given by Eq. \eqref{eq:LineraPotential}. First  we apply a rotation around the $y$ axis \cite{Ibarra-Sierra_2022},
\begin{equation} \label{ec:SolutionOne}
\mathbf{\Psi}(t)
=\exp\left[-i\left(\frac{\pi}{4} 
\right)\sigma_y\right]\mathbf{\Phi}(t).
\end{equation}
Substituting (\ref{ec:SolutionOne}) into 
Eq. (\ref{eq:scrhodinger1}) we obtain,
\begin{equation}
\label{ec:Schrokxky}
i \frac{d}{d\phi}\mathbf{\Phi}(\phi)
=\frac{2}{\hbar \Omega}\left[
\tilde{\Pi}_x\sigma_z
+ \tilde{k}_{y} \sigma_y \right]\mathbf{\mathbf{\Phi}}(\phi) \, ,
\end{equation}
where we defined $\tilde{k}_x=\hbar v_F k_x, \tilde{k}_y=\hbar v_F k_y$
In what follows, we will use a scaled time 
$\phi= \Omega t/2$,
a scaled momentum $\tilde{\Pi}_x=\tilde{k}_x-\zeta_x \cos(2 \phi)$ and a frequency-weighted  induced  dipole  moment,
\begin{equation}
\zeta_x=\frac{ev_xE_x}{\Omega}.
\end{equation}
The spinor components of 
$\mathbf{\Phi}(\phi)=\left(\Phi_{+}(\phi),\Phi_{-}(\phi)\right)^{\top}$
are given by
$\Phi_{+}(\phi)=[\Psi_A(\phi)+\Psi_B(\phi)]/\sqrt{2}$ and 
$\Phi_{-}(\phi)=[\Psi_A(\phi)-\Psi_B(\phi)]/\sqrt{2}$.
A final transformation allows us to remove the term proportional to ${\sigma}_0$ in  Eq. (\ref{ec:Schrokxky})

\begin{equation}\label{ec:suprimdiag}
\mathbf{\Phi}(\phi)=\exp\left[-2i\frac{ \tilde{k}_y}{\hbar \Omega}\, \phi\, \sigma_0\right]\mathbf{\chi}(\phi),
\end{equation}
where $\mathbf{\chi}(\phi)=(\chi_{+1}(\phi),\chi_{-1}(\phi))^{\top}$. 
Finally, if Eq. (\ref{ec:suprimdiag}) is inserted into Eq.
(\ref{ec:Schrokxky}), the following Whittaker-Hill differential equation is found, 
\begin{multline}
\chi_{\eta}''(\phi)+4\bigg[\left(\frac{\tilde{k}_x}{\hbar\Omega}-q_0\cos(2\phi)\right)^{2}+\left(\frac{\tilde{k}_y}{\hbar\Omega}\right)^{2}\\+i\eta q_0\sin(2\phi)\bigg] \chi_{\eta}(\phi)=0.
\label{eq:Almost}
\end{multline}
where $\eta=\pm 1$ is a label for the spinor component and $q_0$ is an important parameter that defines the cases of weak ($q_0>1$) and strong external driving fields ($q_0<1$),
\begin{equation}
    q_0=\frac{\zeta_x}{\hbar \Omega}=\frac{ev_FE_x}{\hbar \Omega^{2}}
\end{equation}
For the circular polarized case, a similar calculation replaces $E_x$ by the norm of the electric field. We remark that in Eq. (\ref{eq:Almost}) the spinors components are completely decoupled. A further advantage is that $\chi_{+1}(\phi)$ and $\chi_{-1}(\phi)$ are the probability amplitudes of the valence and conduction band, respectively. {\color{black} 
Now it is convenient to write Eq. (\ref{eq:Almost}) in the form,
\begin{equation}
\chi_{\eta}''(\phi)+4\bigg[\tilde{\epsilon}^{2}_{\boldsymbol{k}}-q_0g(\phi)
+q_0^{2}f(\phi)\bigg] \chi_{\eta}(\phi)=0.
\label{eq:Whita}
\end{equation}
with,
\begin{equation}
\tilde{\epsilon}^{2}_{\boldsymbol{k}}=\frac{\tilde{k}^{2}_x+\tilde{k}^{2}_y}{(\hbar \Omega)^{2}},
\end{equation}
and two periodic functions defined as,
\begin{equation}
    g(\phi)=\frac{4\tilde{k}_x}{\hbar \Omega}\cos(2\phi)-i\eta 2 \sin(2\phi)
\end{equation}

\begin{equation}
    f(\phi)=2q_0^{2}(1+\cos(4\phi))
\end{equation}

Equation (\ref{eq:Whita}) is a Whittaker-Hill equation—i.e., a harmonic oscillator driven by a periodic force with two harmonics and complex coefficients. It is a relatively unexplored mathematical object, but here we will derive some of its properties.\\

Notice that Eq. (\ref{eq:Whita}) contains several limiting cases depending on the value of $\tilde{\epsilon}_{\boldsymbol{k}}$ when compared with $q_0$. The transition from the strong to the weak case is given by the condition,
\begin{equation}
    \tilde{\epsilon}_{\boldsymbol{k}} \approx q_0
\end{equation}
In Fig. \ref{fig:spectrumhighfield} this transition region is clearly seen in panel b), as it separates regions with quasienergies that follow an elliptic pattern from a region with vertical stripes. As the wave vectors grow in the region $\tilde{\epsilon}_{\boldsymbol{k}} > q_0$ 
, the elliptic pattern becomes circular. This is the weak field region. For  $\tilde{\epsilon}_{\boldsymbol{k}}<q_0$, the pattern is made from vertical lines. Let us explain how these patterns arise.  

The case $k_y=0$ can always be integrated for any field intensity $q_0$ as from Eq. (\ref{eq:Almost}),
\begin{multline}
\chi_{\eta}''(\phi)+4\bigg[\tilde{\epsilon}^{2}_x(\phi)+\frac{i\eta}{2} \frac{d \tilde{\epsilon}_x(\phi)}{d\phi}\bigg] \chi_{\eta}(\phi)=0.
\label{eq:Kxbasiceq}
\end{multline}
where ,
\begin{equation}
    \tilde{\epsilon}_x(\phi)=\left(\frac{\tilde{k}_x}{\hbar\Omega}-q_0\cos(2\phi)\right)
\end{equation}
The solution to this equation is,
\begin{equation}
    \chi_{\eta}(t)= \chi_{\eta}(0)\exp\left[-
    i2\eta  \left( \int_0^{\phi}  \tilde{\epsilon}_x(\phi')d\phi' \right)\right]
\end{equation}
or,
\begin{equation}\label{eq:solcirculos2}
\chi_{\eta}(t)= \chi_{\eta}(0)\exp\left[i\eta\left(q_0 \sin{(\Omega t)}-\frac{\tilde{k}_x t}{\hbar}\right)\right].
\end{equation}
In the previous equation, the phase after one period can be easily determined, from which it follows that the Berry and Aharonov phases are always zero—consistent with Figs. \ref{fig:4} and \ref{fig:6}. Notably, this solution is identical to Eq. (\ref{eq:kxsol}), which was obtained through a different approach.

The general case $k_y \ne 0$ is challenging, as Eq. (\ref{eq:Kxbasiceq}) is no longer valid.
However, for $\tilde{\epsilon}_{\boldsymbol{k}} < q_0$ in Eq. \eqref{eq:Almost}, the term $(\tilde{k}_y/{\hbar \Omega})^{2}$ can be neglected compared to $q_0(\tilde{k}_x/{\hbar \Omega})$. As a result, Eq. \eqref{eq:solcirculos2} closely approximates the solution for a given $\tilde{k}_x$, which explains the vertical stripe pattern observed in the strong field region.

In the weak field region $\tilde{\epsilon}_{\boldsymbol{k}} > q_0$, the elliptic/circular pattern arises from Eq. (\ref{eq:Whita}) as follows. Consider $q_0$ as a perturbation parameter, define $p = d\chi{\eta}(\phi)/d\phi$, and the vector $\mathbf{v} = (\chi_{\eta}(\phi), p)$. The differential equation can be written as,

\begin{equation}
   \frac{d\mathbf{v}}{dt}= \mathcal{A}(t) \mathbf{v}
\end{equation}
with,
\begin{equation}
     \mathcal{A}(t)=\begin{pmatrix}
        0 & 1\\ -4(\tilde{\epsilon}^{2}_{\boldsymbol{k}}-q_0g(\phi)
+q_0^{2}f(\phi)) & 0
    \end{pmatrix}.
\end{equation}
Considering $q_0 \rightarrow 0$ and according to the Floquet theorem, the solutions are determined by the eigenvalues of the matrix $ \mathcal{A}(t)$
\cite{verhulst2006nonlinear}, in this case $\lambda_{\pm}=\pm i 2 \tilde{\epsilon}_{\boldsymbol{k}} $. Taking into account the periodicity of the system we obtain that,
\begin{equation}
   \frac{\sqrt{\tilde{k}_{x}^{2}+\tilde{k}_{y}^{2}}}{\hbar \Omega}=\frac{v_F |\boldsymbol{k}|}{\Omega}=m
\end{equation}
with $m$ an integer. This is the basic structure of the circular pattern seen in the quasienergy Fig. \ref{fig:spectrumhighfield}. The pattern becomes elliptical as $\tilde{\epsilon}_{\boldsymbol{k}} \rightarrow q_0$   since $g(\phi)$ induces an asymmetry due to the term that contains the component $\tilde{k}_x$.
}

\section{Topological phases obtained from the time-evolution operator} \label{Appendix: Approximation-Time-Evolution}

To establish the relationship between the Aharonov-Anandan phase, $\gamma_{A,\boldsymbol{k}}$, and the Berry phase, $\gamma_{B,\boldsymbol{k}}$, we rewrite the Hamiltonian from Eq. \eqref{eq:khamiltonian} for $\xi=1$, in terms of the eigenenergies and instantaneous phases as,
\begin{equation} \label{eq:Hamiltonian-as-function-of-instantaneous-energies}
	{\mathcal{H}}_{\boldsymbol{k}} = |\epsilon_{\boldsymbol{k}}^{\lambda}(t)|\begin{pmatrix}
		0 &  e^{-i \theta_{\boldsymbol{k}}(t)}  \\
		 e^{i \theta_{\boldsymbol{k}}(t)} & 0
	\end{pmatrix}.
\end{equation}
Next, we define the unitary rotation transformation:
\begin{equation} \label{eq:unitary_rotation_transformation}
	\begin{split}
		\mathcal{R}(t) & \equiv \frac{1}{\sqrt{2}} \begin{pmatrix}
			1 & 1 \\
			-e^{i \theta_{\boldsymbol{k}}(t)} & e^{i \theta_{\boldsymbol{k}}(t)}
		\end{pmatrix}, \\
		\mathcal{R}^{\dagger}(t) & \equiv \frac{1}{\sqrt{2}} \begin{pmatrix}
			1 & -e^{-i \theta_{\boldsymbol{k}}(t)} \\
			1 & e^{-i \theta_{\boldsymbol{k}}(t)}
		\end{pmatrix}.
	\end{split}
\end{equation}
This allows us to rewrite the Schrödinger equation in the transformed state as:
\begin{equation} \label{eq:Effective_Hamiltonian}
	i \hbar \frac{d}{dt} \varphi_{\boldsymbol{k}}(t) = \overline{\mathcal{H}}_{\boldsymbol{k}} \varphi_{\boldsymbol{k}}(t)
\end{equation}
Here, the transformed state $\varphi_{\boldsymbol{k}}(t)$ and the effective Hamiltonian $\overline{\mathcal{H}}_{\boldsymbol{k}}$ are defined as:
\begin{equation} \label{eq:definitions_transformation}
	\begin{split}
		\varphi_{\boldsymbol{k}}(t) &\equiv \mathcal{R}^{\dagger}(t) \psi_{\boldsymbol{k}}(t) \\
		\overline{\mathcal{H}}_{\boldsymbol{k}}(t) &\equiv \left\{ \mathcal{R}^{\dagger}(t) {\mathcal{H}}_{\boldsymbol{k}} \mathcal{R}(t) - i \hbar \mathcal{R}^{\dagger}(t) \frac{d}{dt} \mathcal{R}(t) \right\} \\
		&= - \left\{ |\epsilon_{\boldsymbol{k}}^{\lambda}(t)| + \hbar \frac{d \gamma_{B, \boldsymbol{k}}(t)}{dt} \sigma_0 + \hbar \nu_{\boldsymbol{k}}(t) \sigma_x \right\}
	\end{split}
\end{equation}
It can be seen that, with the basis transformation, the Hamiltonian is rewritten in terms of the time derivative of the dynamic, Berry, and Rabi phases.

The unitary time evolution operator can be approximated as:
\begin{equation} \label{eq:Approximation_Unitary_TIme_Evolution}
	\begin{split}
		{\mathcal{U}}(t) &= \exp \left\{ - \frac{i}{\hbar} \int_{0}^{t} \overline{\mathcal{H}}_{\boldsymbol{k}}(t') dt' \right. \\
		&\left. - \frac{1}{\hbar^2} \int_{0}^{t} dt' \int_{0}^{t'} dt'' \left[ \overline{\mathcal{H}}_{\boldsymbol{k}}(t'), \overline{\mathcal{H}}_{\boldsymbol{k}}(t'') \right] \right. \\
		&\left. + \frac{i}{\hbar^3} \int_{0}^{t} dt' \int_{0}^{t'} dt'' \int_{0}^{t''} dt''' \left[ \overline{\mathcal{H}}_{\boldsymbol{k}}(t'), \left[ \overline{\mathcal{H}}_{\boldsymbol{k}}(t''), \overline{\mathcal{H}}_{\boldsymbol{k}}(t''') \right] \right] + \ldots \right\}
	\end{split}
\end{equation}

By evaluating the integrals term by term, we obtain the following contributions:
\begin{equation} \label{eq:integral_term_by_term}
	\begin{split}
		-\frac{i}{\hbar} \int_{0}^{t} \overline{\mathcal{H}}_{\boldsymbol{k}}(t') dt' &= i \left( |\gamma^{\lambda}_{D, \boldsymbol{k}}(t)| \sigma_z \right. \\
		&\left. + \gamma_{B, \boldsymbol{k}} \sigma_0 + \gamma_{R, \boldsymbol{k}} \sigma_x \right) \\
		- \frac{1}{\hbar^2} \int_{0}^{t} dt' \int_{0}^{t'} dt'' \left[ \overline{\mathcal{H}}_{\boldsymbol{k}}(t'), \overline{\mathcal{H}}_{\boldsymbol{k}}(t'') \right] &= 2i \\
		&\times \left\{ \int_{0}^{t} dt' \nu_{\boldsymbol{k}}(t') |\gamma^{\lambda}_{D, \boldsymbol{k}}(t')| \right. \\
		&\left. - \frac{1}{\hbar} \int_{0}^{t} dt' |\epsilon_{\boldsymbol{k}}^{\lambda}(t')| \gamma_{R, \boldsymbol{k}}(t') \right\} \sigma_y
	\end{split}
\end{equation}

For the next term we have:
\begin{equation} \label{eq:integral_term_by_term_2}
	\begin{split}
		\frac{i}{\hbar^3} \int_{0}^{t} dt' \int_{0}^{t'} dt'' \int_{0}^{t''} dt''' \left[ \overline{\mathcal{H}}_{\boldsymbol{k}}(t'), \left[ \overline{\mathcal{H}}_{\boldsymbol{k}}(t''), \overline{\mathcal{H}}_{\boldsymbol{k}}(t''') \right] \right] \\
		= \frac{4i}{\hbar^2} \int_{0}^{t} \int_{0}^{t'} \int_{0}^{t''} dt' dt'' dt''' \left[ \hbar \nu_{\boldsymbol{k}}(t') \sigma_z - |\epsilon^{\lambda}_{\boldsymbol{k}}(t')| \sigma_x \right] \\
		\times \left[ \nu_{\boldsymbol{k}}(t'') |\epsilon^{\lambda}_{\boldsymbol{k}}(t''')| - |\epsilon^{\lambda}_{\boldsymbol{k}}(t'')| \nu_{\boldsymbol{k}}(t''') \right]
	\end{split}
\end{equation}

Finally, the time evolution operator is:
\begin{equation} \label{eq:time_evolution_operator}
	\begin{split}
	{\mathcal{U}}(t) &\approx \exp \left\{ i \left( |\gamma^{\lambda}_{D, \boldsymbol{k}}(t)| \sigma_z + \gamma_{B, \boldsymbol{k}}(t) \sigma_0 + \gamma_{R, \boldsymbol{k}}(t) \sigma_x \right) \right. \\
		&+ 2i \left( \int_{0}^{t} dt' \nu_{\boldsymbol{k}}(t') |\gamma^{\lambda}_{D, \boldsymbol{k}}(t')| - \frac{1}{\hbar} \int_{0}^{t} dt' |\epsilon^{\lambda}_{\boldsymbol{k}}(t')| \gamma_{R, \boldsymbol{k}}(t') \right) \sigma_y \\
		&+ \frac{4i}{\hbar^2} \left( \int_{0}^{t} \int_{0}^{t'} \int_{0}^{t''} dt' dt'' dt''' \left[ \hbar \nu_{\boldsymbol{k}}(t') \sigma_z - |\epsilon^{\lambda}_{\boldsymbol{k}}(t') |\sigma_x \right] \right. \\
		&\times \left. \left[ \nu_{\boldsymbol{k}}(t'') |\epsilon^{\lambda}_{\boldsymbol{k}}(t''')| - |\epsilon^{\lambda}_{\boldsymbol{k}}(t'')|\nu_{\boldsymbol{k}}(t''') \right] \right\}
	\end{split}
\end{equation}

It is important to note that in the adiabatic limit and non-degenerate states, the only phases present are the dynamic phase and the Berry phase, as expected. However, in the non-adiabatic case, correction terms related to transitions between energy levels emerge, particularly the Rabi phase term. These terms are generally incorporated into the Aharonov-Anandan geometric phase discussed in Sec. \ref{sec:topological}.

\newpage

\bibliographystyle{unsrt}

\end{document}